\RequirePackage{lineno} 
\documentclass[aps,prx,twocolumn,floatfix,preprintnumbers,altaffilletter,superscriptaddress,longbibliography]{revtex4-1}
\usepackage{graphicx,url,amssymb,amsmath,rotating,color,units,wasysym,epsfig,multirow,epstopdf,subcaption}
\usepackage[colorlinks,urlcolor=blue,citecolor=blue,linkcolor=blue]{hyperref}
\usepackage{comment}

\newcommand{\lal}{{\sc{LALInference}}}
\newcommand{\xitrue}{4\times10^{-4}}
\newcommand{\nsegs}{17000}
\newcommand{\zmax}{0.77}
\newcommand{\zmin}{0.10}

\newcommand{\dmaxlab}{5}
\newcommand{\dminlab}{0.50}
\usepackage{lineno}
\begin{document}
\title{The optimal search for an astrophysical gravitational-wave background}

\author{Rory Smith}
\email{rory.smith@monash.edu}
\affiliation{Monash Centre for Astrophysics, School of Physics and Astronomy, Monash University, VIC 3800, Australia}
\affiliation{OzGrav: The ARC Centre of Excellence for Gravitational-Wave Discovery, Monash University, VIC 3800, Australia}
\affiliation{LIGO, California Institute of Technology, Pasadena, CA 91125, USA}

\author{Eric Thrane}
\email{eric.thrane@monash.edu}
\affiliation{Monash Centre for Astrophysics, School of Physics and Astronomy, Monash University, VIC 3800, Australia}
\affiliation{OzGrav: The ARC Centre of Excellence for Gravitational-Wave Discovery, Monash University, VIC 3800, Australia}

\begin{abstract}
Roughly every \unit[2-10]{minutes}, a pair of stellar mass black holes merge somewhere in the Universe.
A small fraction of these mergers are detected as individually resolvable gravitational-wave events by advanced detectors such as LIGO and Virgo.
The rest contribute to a stochastic background.
We derive the statistically optimal search strategy (producing minimum credible intervals) for a  background of unresolved binaries.
Our method applies Bayesian parameter estimation to all available data.
Using Monte Carlo simulations, we demonstrate that the search is both ``safe'' and effective: it is not fooled by instrumental artifacts such as glitches and it recovers simulated stochastic signals without bias.
Given realistic assumptions, we estimate that the search can detect the binary black hole background with about one day of design sensitivity data versus $\approx\unit[40]{months}$ using the traditional cross-correlation search.
This framework independently constrains the merger rate and black hole mass distribution, breaking a degeneracy present in the cross-correlation approach.
The search provides a unified framework for population studies of compact binaries, which is cast in terms of hyper-parameter estimation.
We discuss a number of extensions and generalizations including: application to other sources (such as binary neutron stars and continuous-wave sources), simultaneous estimation of a continuous Gaussian background, and applications to pulsar timing.
\end{abstract}

\maketitle

\section{Introduction}
Observations of gravitational waves from binary black hole mergers imply that stellar-mass black holes coalesce somewhere in the visible Universe  every $223^{+352}_{-115}\,$s ~\cite{GW170817_stoch}.
Binary neutron stars merge every $13^{+49}_{-9}\,$s ~\cite{GW170817_stoch}.
The vast majority of these events are too distant to be individually resolved by the current generation of detectors.
The most distant event yet observed, GW170104, was measured to have a redshift of $z=0.18^{+0.08}_{-0.07}$~\cite{GW170104}.
Nonetheless, unresolved compact binary mergers contribute to a stochastic background of gravitational waves, which may be detectable with current detectors~\cite{GW150914_stoch}.
Measuring the stochastic background from compact binaries has the potential to provide information about high-redshift binary black holes and neutron stars, which complements observations of local mergers~\cite{Callister}. 

The stochastic background is typically characterized by the gravitational-wave energy density spectrum
\begin{equation}\label{eq:Omega}
\Omega_\text{gw}(f) \equiv
\frac{1}{\rho_c}
\frac{d\rho_\text{gw}}{d\ln f} ,
\end{equation}
where $d\rho_\text{gw}$ is the energy density of gravitational waves between $f$ and $f + df$ and $\rho_c$ is the critical energy density for a flat universe~\cite{AllenRomano}.
Searches for the stochastic background seek to measure $\Omega_\text{gw}(f)>0$.
In the LIGO/Virgo band ($\unit[10-2000]{Hz}$), the best current limits on the gravitational-wave energy density spectrum are $\Omega_\text{gw}(f)<1.7\times10^{-7}$ (95\% confidence, measured in the band $\unit[20-86]{Hz}$)~\cite{O1_iso,O1_dir}.

To date, all LIGO/Virgo searches for the stochastic background have relied on the cross-correlation method described by Allen \& Romano~\cite{AllenRomano}.
A similar cross-correlation technique is employed by pulsar timing arrays operating in the nanohertz band~\cite{HellingsDowns}.
The cross-correlation method has two nice features.
It is computationally cheap and it yields a statistically optimal (minimum-variance) measurement of $\Omega_\text{gw}(f)$ for the case of a persistent, {\em Gaussian} background.

A Gaussian background is characterized entirely by $\Omega_\text{gw}(f)$.
The stochastic background from stellar-mass binary black holes is highly {\em non-Gaussian}; it is rare for different events to overlap in time \cite{GW170817_stoch} (within the advanced detector band).
The stochastic backgrounds from binary neutron stars are quasi-Gaussian since the signals from individual events overlap in time~\cite{GW170817_stoch}.
We focus initially on the highly non-Gaussian background from stellar-mass binary black holes, but return below to consider quasi-Gaussian backgrounds from binary neutron stars.

The binary black hole background consists of (in-principle), clearly-distinguishable, deterministic signals~\cite{Rosado}.
Since distant events are not resolvable with current detectors, most events are not distinguishable {\em in practice}, but they would be with a more sensitive detector. This is in contrast to, for example, the stochastic background from white dwarf binaries in the millihertz band, which cannot be distinguished {\em in principle}; see, e.g.,~\cite{Cutler98}.
Cross-correlation searches are sub-optimal for non-Gaussian backgrounds.
It is possible to improve the sensitivity of the stochastic search by including a more accurate description of the signal model.

A number of studies dating back to 2002 have outlined a variety of strategies for developing a non-Gaussian pipeline.
Drasco \& Flanagan derived an algorithm suitable for bursting sources~\footnote{A practical working definition of ``bursting'' sources is transient signals that are not compact binary coalescences and which are not modeled with a bank of templates.} observed by co-located detectors with white noise~\cite{DrascoFlanagan}.
While the assumptions of white noise and co-located detectors are not realistic, the study was important for showing that a non-Gaussian analysis could achieve a potentially significant improvement in sensitivity compared to a cross-correlation search.
Subsequently, Thrane outlined a method for bursting sources that can be applied in the more realistic case of spatially separated detectors with colored noise~\cite{popcorn}.
Work by Martellini \& Regimbau has explored additional strategies~\cite{Martellini14,Martellini15}.
None of these methods have yet been used in a published search.
For a comprehensive review of stochastic background methodology up until this point, see~\cite{RomanoCornish}.

In this paper, we take a step back and ask a new question: what is the {\em optimal} method for detecting a stochastic background of binary black holes (or any other astrophysical background)?
Formally, the optimal method yields a minimum credible interval posterior.
In practical terms, the optimal method can be more sensitive than sub-optimal techniques.
Conceptually, deriving the optimal method amounts to implementing a likelihood function, which describes the salient features of the signal model and measurement noise.
Turning the crank of Bayesian statistics, the resulting posterior distributions yield minimum credible intervals by construction.

It turns out that this question of optimality is interesting for several reasons.
First and most obvious, it is desirable to derive the most sensitive possible search.
Depending on the magnitude of improvement over the cross-correlation search, an optimal pipeline could significantly reduce the time to the detection of a stochastic background.
Second, we show below that the optimal Bayesian search yields, as a byproduct, information about the population properties of high-redshift black holes.
In particular, we automatically obtain a posterior distribution for the coalescence rate of binary black holes at high redshift.
This information is lost during the process of cross-correlation.

Third, our formalism provides a natural framework for unifying the stochastic search with measurements of unambiguous detections and even ``silver-plated" candidates~\cite{Cannon13} like LVT151012~\cite{O1_BBH}.
As the number of detected black hole mergers increases, the gravitational-wave community is likely to be increasingly interested in population statements.
The framework proposed here is the natural means of combining all possible data to make statements about populations of binary black holes (and other astrophysical phenomena).
The method is free of Malmquist bias~\cite{malmquist} since there is no selection of events.

Last, we outline how the method can be generalized in order to solve a number of important problems in gravitational-wave astronomy including (1) measurement of a primordial Gaussian background in the presence of an astrophysical foreground and (2) measurement of the population properties of binary black holes and neutron stars, e.g., their mass and spin distributions.

The remainder of this paper is organized as follows.
In Section~\ref{formalism}, we derive the optimal search for an ensemble of binary black hole mergers.
In Section~\ref{mdc}, we present the results of a Monte Carlo study that demonstrates the method.
In Section~\ref{scaling}, we investigate scaling relations in order to estimate the performance of the search in various contexts.
In Section~\ref{noise}, we consider complications arising from non-Gaussian noise.
In Section~\ref{computing}, we consider the feasibility of the search given plausible computing resources.

In Section~\ref{hyper}, we discuss how this method can be used to measure the population properties of unresolved binary black holes using hyper-parameters.
In Sec.~\ref{extensions}, we discuss how the method can be generalized to similar problems including binary neutron stars, neutron star black hole binaries, continuous waves from rotating neutron stars, and supermassive black hole binaries.
In Section~\ref{Gaussian}, we discuss the simultaneous measurement of a primordial gravitational-wave background.
In Section~\ref{conclusions}, we provide an overall assessment of the feasibility of an optimal stochastic search and the prospects for detection of a stochastic background.

\section{Method}\label{formalism}
\subsection{The likelihood function}
We break the strain data $s$ into convenient-sized segments.
The data for segment $i$ is denoted $s_i$.
By assumption, each segment is big enough that it might include one signal (specifically, the merger part of the signal), but small enough that it is unlikely to contain two such signals.
In this paper, we satisfy these criteria by using segments with a duration of $\unit[4]{s}$.
For binary black holes like GW150914, the signal is in-band for only $\approx\unit[0.2]{s}$, less than the segment duration.
Since the probability of observing just one event is in any given segment is small $\approx2\%$, the probability of observing two at once is negligibly small: $\approx10^{-4}$.

Assuming Gaussian noise, the log likelihood for a single segment $s_i$ containing a compact binary signal with parameters $\theta$ is
\begin{equation}\label{eq:L}
\log\big[{\cal L}(s_i|\theta_i)\big]
    \propto -\frac{1}{2}\big\langle s_i-h(\theta_i), s_i-h(\theta_i) \big\rangle
\end{equation}
Here $h(\theta_i)$ is the signal model, which depends on a vector of parameters $\theta_i$, e.g., sky location, component masses, the time of coalescence within the segment, and so on.
We have introduced the usual noise-weighted inner product:
\begin{equation}\label{eq:inner}
    \big\langle a, b \big\rangle \equiv 4\Re\Delta f \sum_{k}\frac{a^{*}(f_k)\,b(f_k)}{S_n(f_k)} \,,
\end{equation}
where $\Delta f$ is the frequency-bin size and $\Re$ is the real part of the sum.
The variable $S_n(f_k)$ is the detector noise power spectral density.
The sum runs over $k$ frequency bins.

When combining data from a network of $M$ detectors, the likelihood function for segment $i$ becomes
\begin{equation}
\label{eq:multidetectorL}
{\cal L}(\vec{s}_i|\theta) = \prod_{j=1}^M {\cal L}(s_{i}^{(j)}|\theta_i) ,
\end{equation}
where $\vec{s}_i$ represents the data from all $M$ detectors and $j$ indexes each of the M detectors.
Stochastic searches typically rely on $M\geq2$ networks in order to distinguish astrophysical signals from poorly modeled noise~\cite{AllenRomano}.
Henceforth, we assume $M\geq2$ detectors because this enables a \textit{coherent} search for sub-threshold signals.   
The notation $\vec{s}_i$ is used to indicate the strain data from $M$ detectors associated with segment $i$. 

We can generalize the likelihood function in  Eq.~\ref{eq:L} to account for the fact that we  expect the data to contain a signal plus Gaussian noise with probability $\xi$, \textit{or} pure Gaussian noise with probability $(1-\xi)$.
(We defer discussion of non-Gaussian noise until Section~\ref{noise}.)
Hence, we define a ``generalized likelihood'' for segment $i$, modified to include the signal probability hyper-parameter $\xi$~\cite{DrascoFlanagan}, which we refer to as the ``duty cycle'' 
\begin{align}\label{eq:GenLike}
\mathfrak{L}(\vec{s}_i|\theta_i,\xi) =
\xi \, {\cal L}(\vec{s}_i|\theta_i)
+ (1-\xi) \, {\cal L}(\vec{s}_i|0) .
\end{align}
We note that this astrophysical duty cycle should not be confused with the detector duty cycle, which is the fraction of time during which a detector collects science-quality data.
Here, ${\cal L}(\vec{s}_i|0)$ is the likelihood function given the hypothesis that no signal is present:
\begin{equation}
\log\big[{\cal L}(\vec{s}_i|0)\big] \propto -\frac{1}{2}\big\langle s_i, s_i \big\rangle .
\end{equation}

We marginalize over the astrophysical parameters of the event $\theta_i$ (with a prior distribution $\pi(\theta_i)$) in order to obtain a likelihood for the data given the duty cycle $\xi$ for segment $i$
\begin{align}\label{eq:marginalized}
\mathfrak{L}(\vec{s}_i|\xi) 
= \xi \, {\cal Z}_S^i + 
(1-\xi) {\cal Z}_N^i .
\end{align}
We have introduced two new terms
\begin{align}
{\cal Z}_S^i \equiv & \int d\theta \, 
{\cal L}(\vec{s}_i|\theta_i) \, 
\pi(\theta_i) \label{eq:ZS}\\
{\cal Z}_N^i \equiv & {\cal L}(\vec{s}_i|0) \label{eq:ZN} ,
\end{align}
corresponding respectively to the signal evidence and the noise evidence.
These evidences are the raw output of LIGO parameter estimation algorithms such as \lal\ \cite{PhysRevD.91.042003}.
We do not marginalize over the detector noise power spectral density.
The power spectral densities that enter Eq.~\ref{eq:inner} are empirically estimated for each data segment.

The next step is to combine data from a large set of $n$ segments $\{\vec{s}\}$.
The combined likelihood for the data given $\xi$ is 
\begin{align}\label{eq:product}
\mathfrak{L}^\text{tot}\big(\{\vec{s}\}|\xi\big) = & \prod_i^n \mathfrak{L}(\vec{s}_i|\xi) \nonumber\\
= & \prod_i^n \big( \xi \, {\cal Z}_S^i + 
(1-\xi) {\cal Z}_N^i \big).
\end{align}

The posterior for the duty cycle $p(\xi|\{\vec{s}\})$ is  simply
\begin{equation}\label{eq:p_xi}
p(\xi|\{\vec{s}\}) \propto \mathfrak{L}^\text{tot}\big(\{\vec{s}\}|\xi\big)\pi(\xi)\,,
\end{equation}
where $\pi(\xi)$ is the prior distribution on $\xi$.
For simplicity we will assume a flat, uniform prior for $\pi(\xi)$ so that 
\begin{align}
p(\xi|\{\vec{s}\}) \propto \mathfrak{L}^\text{tot}\big(\{\vec{s}\}|\xi\big) ,
\end{align}
though our analysis can incorporate any suitable choice of $\pi(\xi)$, informed, for instance, by expectations about the average time between binary black hole mergers \cite{GW170817_stoch}. 

\subsection{Detection statistic}
In order to search for an astrophysical stochastic background, we calculate a ``stochastic background evidence''
\begin{equation}
{\cal Z}_\text{stoch} = \int d\xi \, {\cal L}\big(\{\vec{s}\}|\xi\big) \pi(\xi) .
\end{equation}
The null hypothesis (there is no stochastic background) is described by a null evidence:
\begin{equation}
{\cal Z}_0 = {\cal L}\big(\{\vec{s}\} | \xi=0 \big)
\end{equation}
We construct a Bayes factor to compare the two hypotheses:
\begin{equation}\label{eq:bf}
\text{BF} = {\cal Z}_\text{stoch} / {\cal Z}_0 .
\end{equation}
This variable is an optimal detection statistic for an astrophysical background of compact binaries. The optimality follows from the fact that the likelihood function describing the data given the signal and noise models is complete. Hence, the search produces a minimum credible-interval posterior distribution on the duty cycle.
In the remainder of this paper, we adopt the convention that a log Bayes factor of $\approx8$ represents a statistically significant preference for one hypothesis over the other~\cite{Jeffreys61}.

Up to this point, we have, for the sake of convenience, written our likelihoods in terms of duty cycle $\xi$.
In the subsequent subsections (\ref{rate}-\ref{Omega}), we discuss how $\xi$ is related to other quantities including rate and energy density.

\subsection{Rate}\label{rate}
In this subsection, we take the likelihood $\mathfrak{L}^\text{tot}\big(\{\vec{s}\}|\xi\big)$---a function of duty cycle $\xi$---and recast it as a function of $R$: the number of mergers per segment.
In particular, $R$ is the rate of events {\em throughout the visible Universe} per segment.
This will allow us to more easily relate our analysis to observations of individual merger events.
It is useful to contrast $\xi$ and $R$.
While duty cycle $\xi$ is defined on $(0,1)$, the rate of events per unit segment is defined on $(0,\infty)$.
Even perfect knowledge of $\xi$ does not determine $R$ because the latter is subject to cosmic variance arising from the fact that events take place randomly following Poisson statistics.

The number of compact binary mergers $N$ is given by the product of the duty cycle $\xi$ and the number of segments $n$
\begin{equation}
N = n\,\xi .
\end{equation}
We perform a change of variable in order to recast the likelihood variable in terms of $N$:
\begin{align}
\mathfrak{L}^\text{tot}\big(\{\vec{s}\}|N\big) = \int d\xi \,
\mathfrak{L}^\text{tot}\big(\{\vec{s}\}|\xi\big) \, 
\pi(\xi | N) ,
\end{align}
where 
\begin{align}
\pi(\xi | N) = 
\left|\frac{dN}{d\xi}\right| \pi(N)
= n \, \pi(N) .
\end{align}

Having recast the likelihood in terms of the number of compact binary mergers $N$, we can marginalize over $N$ to obtain
\begin{align}\label{eq:posterior_N}
{\cal L}^\text{tot}(\{\vec{s}\}|R) = \sum_N
\mathfrak{L}^\text{tot}\big(\{\vec{s}\}|N\big) \, \pi(N|R) ,
\end{align}
where $\pi(N|R)$ is a conditional prior for the number of compact binary mergers $N$ given a rate $R$ (with units of mergers per segment).
The conditional prior is given by a Poisson distribution
\begin{equation}
\pi(N|R) = e^{-R}\frac{R^N}{N!} .
\end{equation}
The total evidence is a likelihood function for the data given the rate hyper-parameter
\begin{align}
\mathfrak{L}^\text{tot}\big(\{\vec{s}\}|R\big) = & {\cal Z}^\text{tot}(R) \\
= & \sum_N
\mathfrak{L}^\text{tot}\big(\{\vec{s}\}|N\big) \, e^{-R}\frac{R^N}{N!} .
\end{align}
The rate posterior is 
\begin{equation}
p(R|\{\vec{s}\}) \propto \mathfrak{L}^\text{tot}(\{\vec{s}\}|R) \, \pi(R) ,
\end{equation}
where $\pi(R)$ is some suitable prior on the rate $R$.

\subsection{Local rate}
The variable that we have been referring to as ``the rate'' $R$---by which we mean, the number of compact binary mergers in the visible Universe per segment---is distinct from {\em the local merger rate} $R_0$ with units of $\unit[]{Gpc^{-3}\,yr^{-1}}$.
However, they are related. We derive the local rate $R_0$ from $R$:
\begin{equation}\label{eq:R(z)}
\frac{R}{\delta t} = \int \frac{dz}{1+z} \left(\frac{dV}{dz}\right) R(z) ,
\end{equation}
Here, $R(z)$ is the co-moving merger rate in units of $\unit[]{Gpc^{-3}\,yr^{-1}}$ as a function of redshift and $\delta t$ is the segment duration.
The factor $dV/dz$ describes how an element of volume evolves in an expanding universe while the factor of $1+z$ comes about transforming the time variable in the source frame to the detector frame; see~\cite{GW170817_stoch}.

The shape of $R(z)$ is determined by a model, which takes into account, e.g., the stellar formation rate as a function of redshift and the time delay between formation and coalescence; see, e.g.,~\cite{GW150914_stoch}.
However, we can treat the overall normalization as a free parameter so that
\begin{equation}\label{eq:shape}
R(z) \equiv R_0 \, \mathfrak{S}(z) ,
\end{equation}
where $\mathfrak{S}(z)$ is a model-dependent, dimensionless shape function (normalized so that $\mathfrak{S}(z=0)=1$) and the local rate $R_0\equiv R(z=0)$ is the normalization.
Combining Eqs.~\ref{eq:R(z)} and~\ref{eq:shape}, we obtain
\begin{equation}
R_0 = \left. \frac{R}{\delta t} \middle / \int \frac{dz}{1+z} \left(\frac{dV}{dz}\right) \mathfrak{S}(z) \right. .
\end{equation}
It will be interesting to compare the local rate inferred from a stochastic detection---and assuming some shape model $\mathfrak{S}(z)$---with the local rate measured directly by resolvable mergers.
Tension in these two measurements could indicate an inadequate shape model $\mathfrak{S}(z)$ among other things.

\subsection{Energy density}\label{Omega}
Given some model, the local rate $R_0$ (and therefore $R$) can be converted into dimensionless energy density $\Omega_\text{gw}(f)$ defined in Eq.~\ref{eq:Omega}; see, e.g.,~\cite{XJ,2001astro.ph..8028P}.
The fully general expression is a bit unwieldy so we employ two simplifying assumptions in order to obtain an intuitive initial expression.
First, we assume that the (source-frame) energy spectrum of the event $dE_\text{GW}/df_s$---a function of the source frame frequency $f_s$---is determined primarily by the chirp mass of the binary $M_c$.
Second, we assume that the co-moving merger rate does not depend on mass.
Given these two assumptions,
\begin{align}\label{eq:Omega2}
\Omega_\text{gw}(f) = & \left(\frac{f}{\rho_c}\right)
\left(\int_0^\infty \frac{dz}{1+z} \frac{R_0 \, \mathfrak{S}(z) }{H(z)} \right) \nonumber\\
& \left( \int dM_c \frac{dE_\text{gw}}{df_s}(f_s|M_c) \pi(M_c)\right) ,
\end{align}
where
\begin{align}\label{eq:E}
H(z) = & H_0 \sqrt{\Omega_M(1+z)^3 + \Omega_\Lambda } .
\end{align}
$H_0$ is the Hubble constant, $\Omega_M$ is the energy density of matter, and $\Omega_\Lambda$ is the energy density of the cosmological constant.
In the first set of parentheses, $\rho_c$ is the critical energy density for a flat universe.
The variable $\pi(M_c)$ is the mass distribution.

Thus, we can relate the number of events per segment $R$ (and/or the local rate $R_0$)  to the energy density spectrum $\Omega_\text{gw}(f)$, but only by employing a model to describe the distribution of events in redshift, mass, and so on.
In the derivation above, we assumed that $\Omega_\text{gw}(f)$ depends primarily on chirp mass.
However, a more general expression for $\Omega_\text{gw}(f)$ would include integrals over every variable that can affect the energy spectrum, for example, the mass ratio.
Also, in general, $\mathfrak{S}(z)$ can depend on variables such as $M_c$, in which case $\mathfrak{S}(z|M_c,...)$ cannot be taken out of the Eq.~\ref{eq:Omega2} integral over $M_c$.

\section{Demonstration Using Gaussian Noise
}\label{mdc}
In this section we carry out a Monte Carlo simulation in order to demonstrate the method described in Section~\ref{formalism}.
Our goal is to calculate the duty cycle posterior $p(\xi|\{\vec{s}\})$ using simulated data containing a population of sub-threshold black hole binaries.
We seek to fulfill three criteria.
First, the method should be ``safe'': it should return a null result when applied to pure Gaussian noise.
Second, the method should be effective: it must yield a positive detection in the presence of a sufficiently loud stochastic signal.
Third, the method should be unbiased.
The duty cycle posterior ought to, on average, peak at the injected value.

We assume a two-detector network consisting of the LIGO Hanford and Livingston observatories operating at design sensitivity~\cite{aligo}.
It is straightforward to extend the method to include additional detectors \footnote{To include additional detectors, one simply uses the expression for the likelihood function in Eq.~\ref{eq:multidetectorL} setting $M$ to the required number of detectors.}, but we begin with two for the sake of simplicity.
For each detector, we generate two datasets.
The {\em noise} dataset consists of 1000 $\unit[4]{s}$ segments of Gaussian noise.
The {\em signal} dataset consists of 300 $\unit[4]{s}$ segments with a binary black hole signal added to Gaussian noise.

The signals are coherently generated (and later recovered) using the \texttt{IMRPhenomPv2} approximant~\cite{IMRPhenomP}.
The parameters of each merger are drawn randomly.
The orientation angles and sky position are drawn from isotropic distributions.
We marginalize over the time of coalescence, which is drawn from a uniform distribution.
The total mass is drawn from a uniform distribution on $(48 M_\odot, 80 M_\odot)$ while the mass ratio is drawn from a uniform distribution on $(1,8)$.
This mass range produces signals that fit conveniently into our $\unit[4]{s}$ segments, given a minimum frequency of $\unit[20]{Hz}$.
The mass ratio is within the domain of validity for \texttt{IMRPhenomPv2}.
The dimensionless spin magnitudes $(a_1, a_2)$ are drawn from a uniform distribution on $(0,0.89)$, which is within the domain of validity for \texttt{IMRPhenomPv2}.
The spin unit vectors are drawn from an isotropic distribution. 
The luminosity distances are drawn from a uniform-in-volume distribution on the interval $(\unit[\dminlab]{Gpc},\unit[\dmaxlab]{Gpc})$.
The lower limit of $\unit[\dminlab]{Gpc}$ removes most gold-plated detections, the remainder of which are eliminated with an additional cut described below.

Reconstructed masses and distance are affected by redshift.
The {\em lab-frame} masses $m_{i=1,2}^l$ and the luminosity distance $d_L$ are given by
\begin{align}
m_i^l = & (1+z) m_i^s \\
d_L = & (1+z) d_M
\end{align}
where $m_i^s$ is measured in the {\em source frame} and $d_M$ is the comoving distance.
The mass distributions described above apply to quantities measured in the lab frame.

Luminosity distances of $(\unit[\dminlab]{Gpc},\unit[\dmaxlab]{Gpc})$ imply a redshift interval of $(\zmin,\zmax)$ within the framework of the standard $\Lambda$CDM cosmology.
In principle, we can (and eventually will) extend the interval to $d\gtrsim\unit[30]{Gpc}$ ($z\gtrsim3.3$) beyond which the stochastic signal is expected to be marginal due to the low stellar formation rate and expanding Universe~\cite{Callister}.
The uniform-in-volume distribution is then modified by the imposition of a ``Malmquist prior''~\cite{malmquist}; we exclude ``gold-plated'' detections that can be observed with a statistically significant coherent matched-filter signal-to-noise ratio $\rho_\text{network}\geq12$:
\begin{equation}\label{eq:rho}
\rho_{\text{network}} = \max_\theta
\sqrt{ \sum_{j} \frac{\big\langle s_j, h(\theta)\big\rangle^{2}}{ \big\langle h(\theta), h(\theta)\big\rangle} }\,.
\end{equation}
Here, the sum over $j$ runs over detectors.
The maximum over $\theta$ determines the matched filter template that best fits the data, and so $\rho_\text{network}$ is the maximum-likelihood signal-to-noise ratio.
Since most binaries merge near the edge of our $\unit[\dmaxlab]{Gpc}$ sphere, this cut removes $\lesssim1\%$ of the events.

\begin{figure}[h]
\includegraphics[width=\columnwidth]{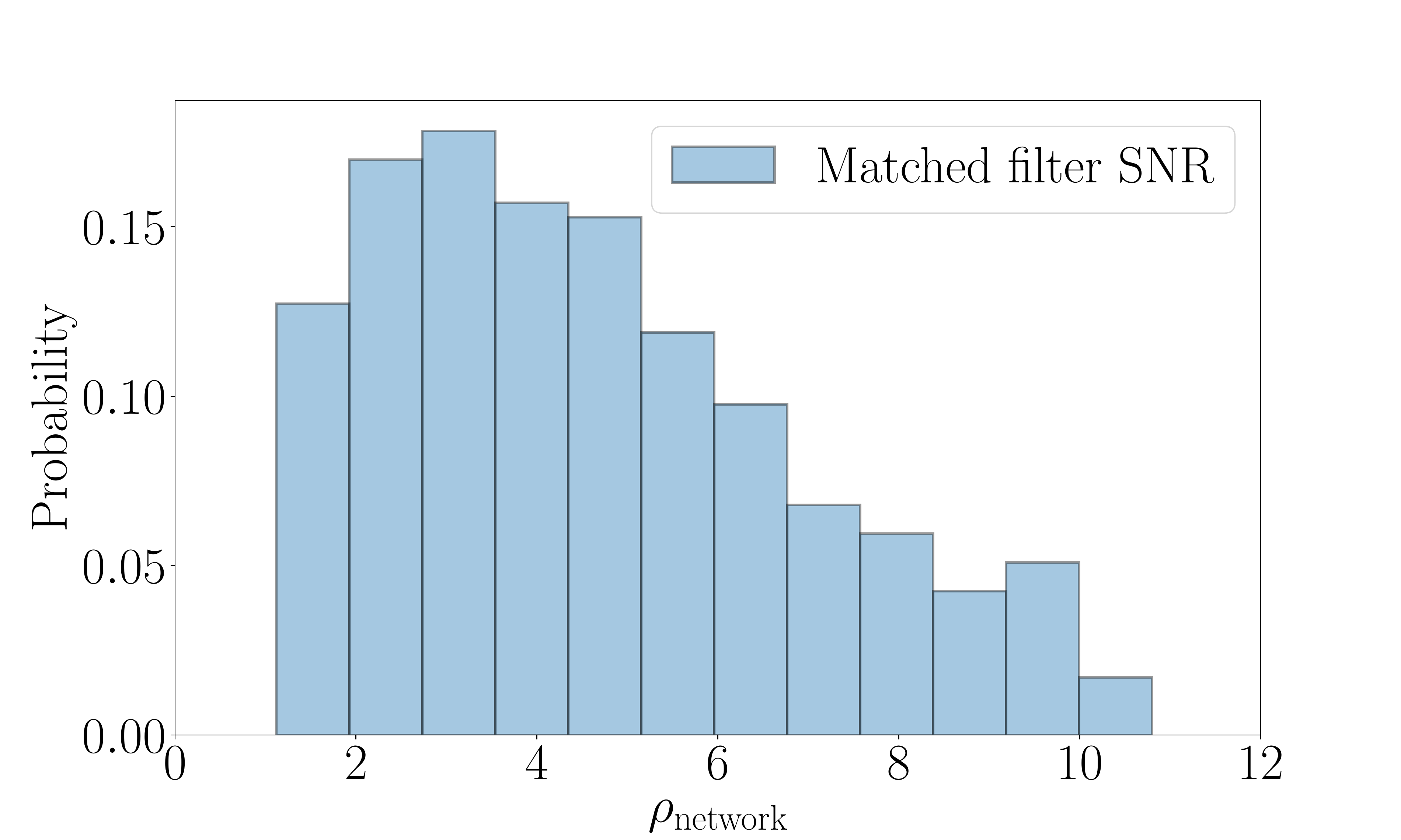}
\caption{
Histogram of the coherent network matched filter signal-to-noise ratio  (Eq.~\ref{eq:rho}) for a dataset containing simulated signals + Gaussian noise.
The network signal-to-noise ratio is calculated by maximizing over all astrophysical parameters.
This is why the distribution falls off sharply near $\rho_\text{network}=1$; one of the many available templates tends to produce a reasonable fit to the data  such that $\rho_\text{network}\gtrsim1$, even for weak signals.
The signal distribution is generated from a population of mergers uniform in volume between $(\unit[\dminlab]{Gpc},\unit[\dmaxlab]{Gpc})$.
Gold-plated detections with $\rho_{\text{network}}\geq12$ are excluded.
}
\label{fig:snr_dist}
\end{figure}

Eventually, gold-plated events should be included in the analysis in order to achieve a unified approach to compact binary population inference and stochastic backgrounds.
However, we exclude them here for the sole purpose of demonstrating that we can recover the signal from a stochastic background of sub-threshold events.
A histogram of $\rho_{\text{network}}$ is shown in Fig.~\ref{fig:snr_dist}.

Having generated mock data for the noise and signal populations, we create mixed populations for arbitrary duty cycles $\xi$ by selecting at random a mixture of entries from each distribution.
We construct three datasets corresponding to $\xi=(0,0.05,1)$ with $n=(500,525,300)$ segments respectively.
We compute the signal evidence ${\cal Z}_S$ and noise evidence ${\cal Z}_N$ for every event in each dataset (Eqs.~\ref{eq:ZS} and~\ref{eq:ZN}).
The calculation is carried out using \lal.
We employ reduced order modeling and reduced order quadrature methods to control the computational cost of the analysis \cite{PhysRevD.94.044031}.

\begin{figure}
        \centering
        \begin{subfigure}[t]{\columnwidth}       \includegraphics[width=\columnwidth]{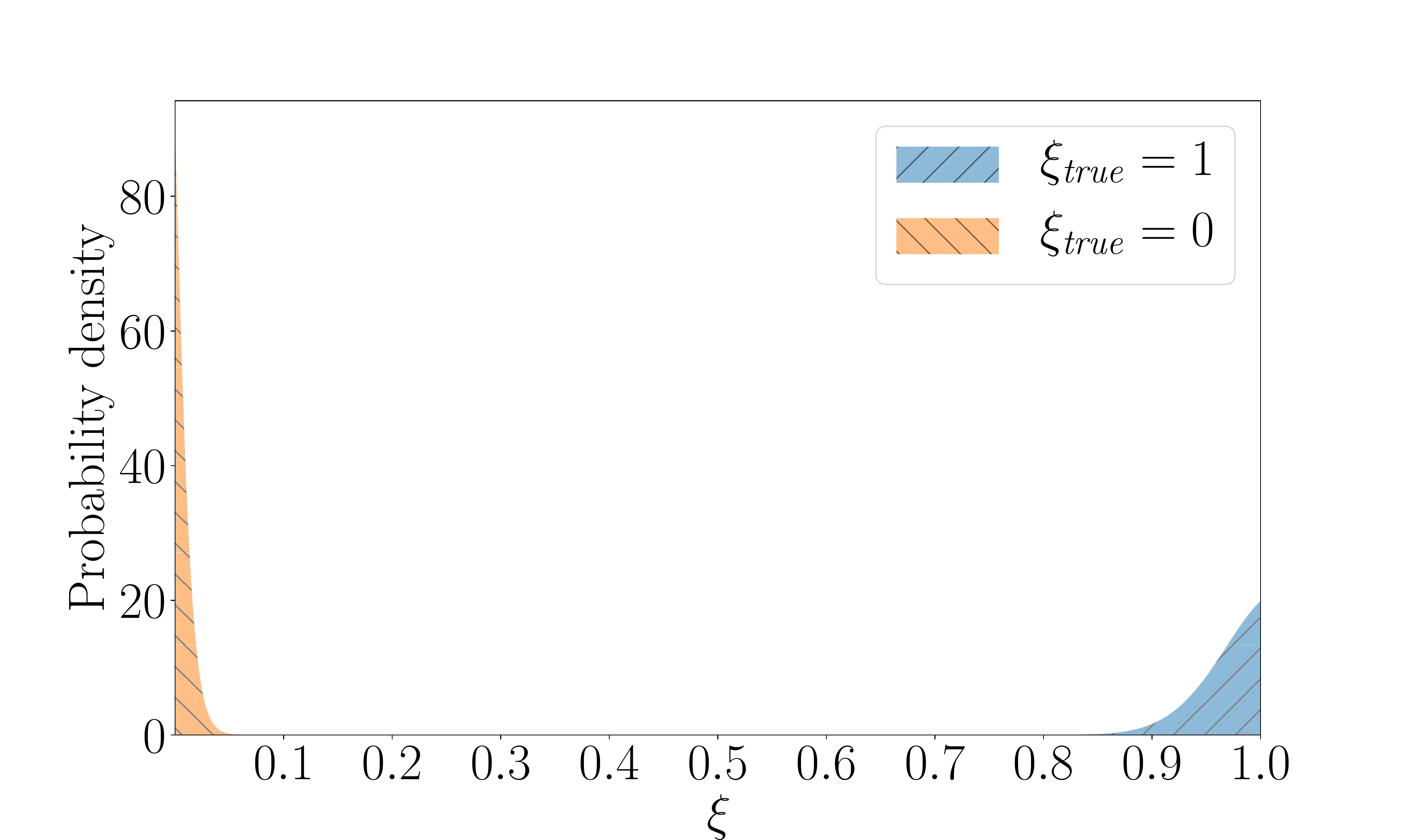}
                \caption{
                Safe and effective duty cycle posteriors.
                }
        \end{subfigure}
        \begin{subfigure}[t]{\columnwidth}  \includegraphics[width=\columnwidth]{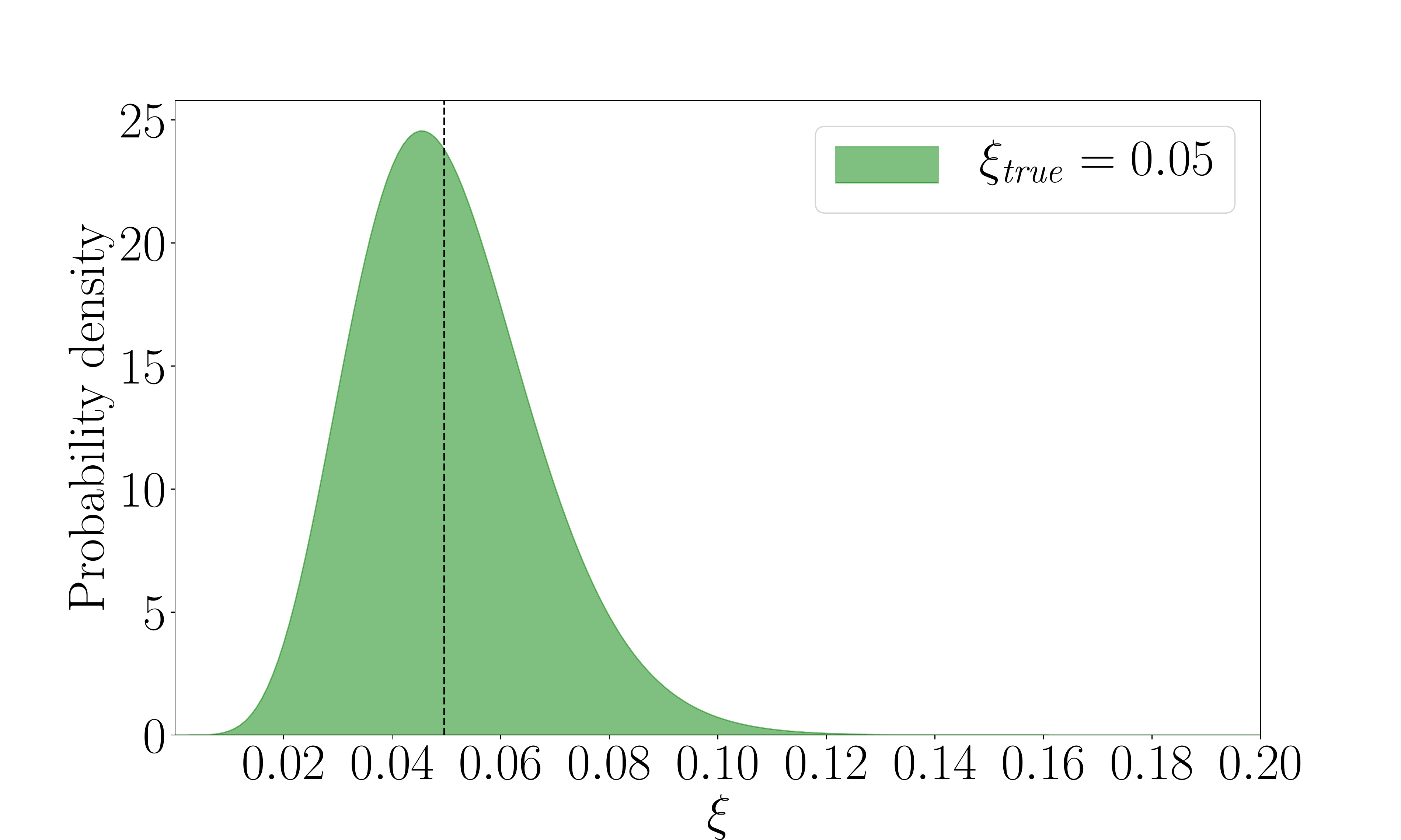}
                \caption{
                Unbiased duty cycle posterior for Gaussian noise containing software injections with a duty cycle $\xi_{\mathrm{true}}=0.05$. The vertical line shows the true duty cycle.
                }
        \end{subfigure}
        \caption{
        Duty cycle posteriors $p(\xi|h)$ (Eq.~\ref{eq:p_xi}) for Monte Carlo datasets.
The orange data are pure Gaussian noise, the blue data are a population of sub-threshold binary black hole events added to Gaussian noise, and the green data correspond to a mixture with a true duty cycle of $\xi_\text{true}=0.05$.
The fact that each posterior peaks at the appropriate value shows that the method is safe, effective, and unbiased. 
        }
\label{fig:gaussian_pxi}
\end{figure}

In Fig.~\ref{fig:gaussian_pxi}, we plot $p(\xi|\{\vec{s}\})$ for the three values of duty cycle.
The posterior for pure noise ($\xi_{\text{true}}=0$) is indicated by $\backslash$-hatched orange.
The fact that it peaks at $\xi=0$ shows that the method is safe.
The posterior for pure signal ($\xi_{\text{true}}=1$) is indicated by $\slash$-hatched blue.
The fact that it peaks at $\xi=1$, clearly excluding $\xi=0$, indicates that the method is effective.
Finally, the posterior for a mixed distribution ($\xi_{\text{true}}=0.05$) is indicated by green.
The posterior peaks near the true value of $\xi_\text{true}=0.05$, which shows that the method is unbiased.

\section{Time to detection}\label{scaling}
In this section, we estimate the time it will take to make a confident detection of an astrophysical background ($\ln\text{BF}>8$).
In doing so, we study how the Bayes factor (Eq.~\ref{eq:bf}) scales as a function of the true duty cycle $\xi_{\text{true}}$ and the number of data segments $N_\text{segs}$.
Recent estimates place the binary black hole background at $\Omega_\text{gw}(f=\unit[25]{Hz})\approx 1.1\times10^{-9}$~\cite{GW170817_stoch}.
In Appendix~\ref{xi_derivation}, we show that, given our mass and distance distributions (described above in Section~\ref{mdc}), $\xi_\text{true}=\xitrue$ provides a realistic duty cycle.
We assume this value of $\xi_\text{true}$ for the remainder of this section.
Due to computational constraints, we create $1000$ noise segments for this preliminary study.
This allows us to carry out a mock study using $\approx\unit[1]{hour}$ of data and to probe duty cycles $\xi\gtrsim0.1\%$.
In order to estimate how the algorithm  will perform when applied to longer datasets and/or lower duty cycles, it is necessary to extrapolate.
Our extrapolation provides an initial performance estimate, which should be checked with a more computationally expensive mock data challenge.

\begin{figure*}[htb] 
  \begin{subfigure}[c]{0.49\linewidth}
    \includegraphics[width=1\linewidth]{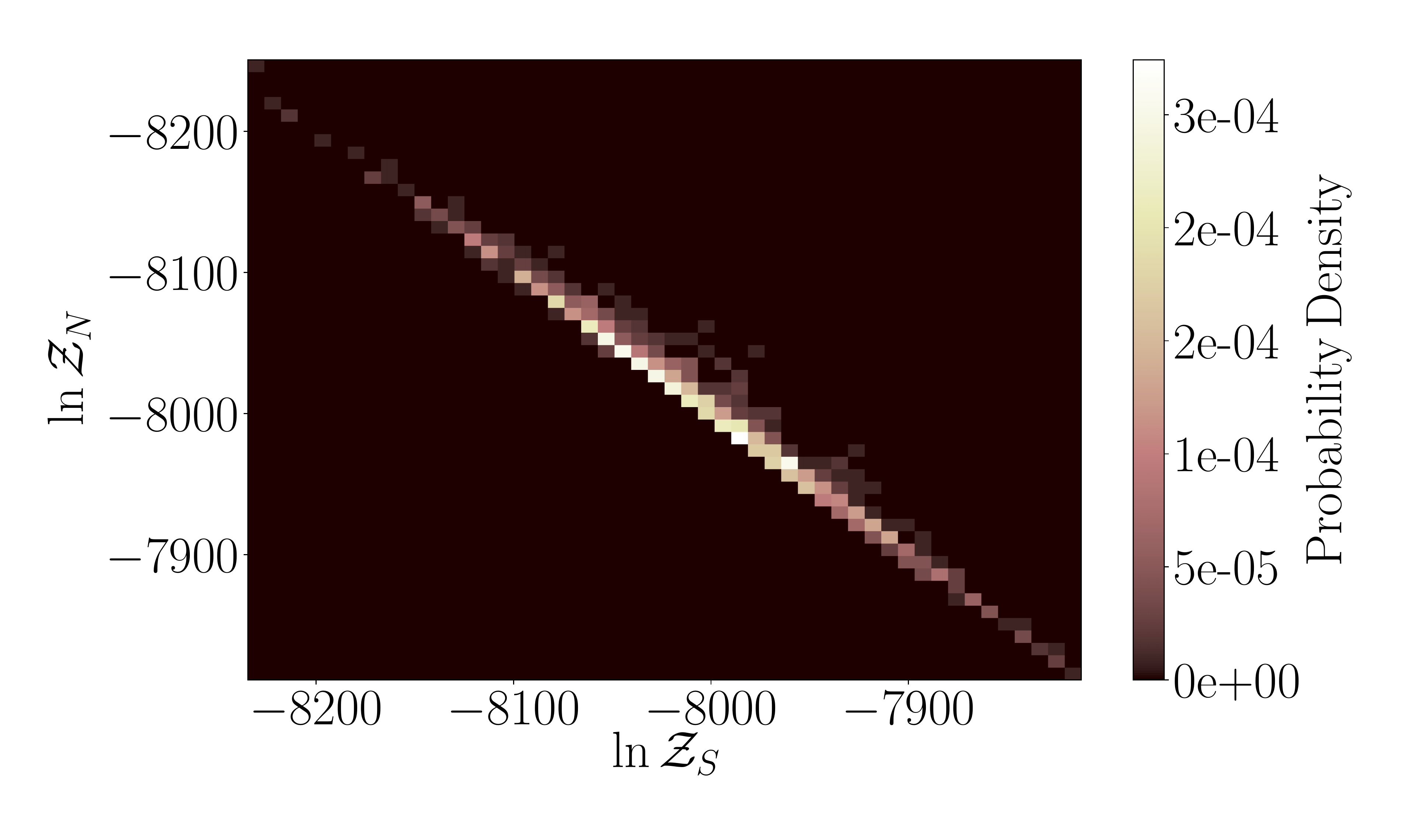} 
  \end{subfigure}
  \begin{subfigure}[c]{0.49\linewidth}
    \includegraphics[width=1\linewidth]{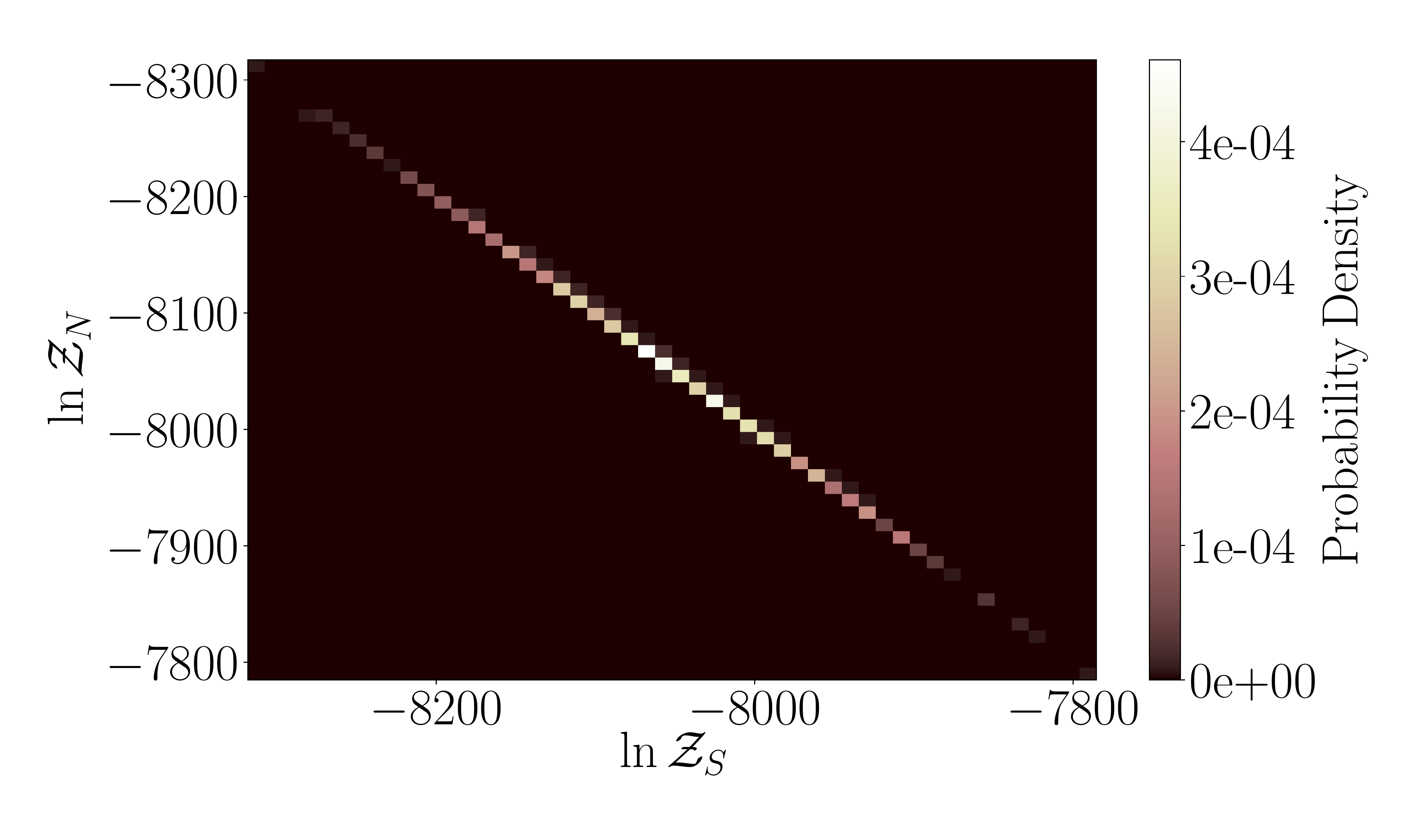} 
  \end{subfigure}
      \caption{
    Probability density functions for signal and noise evidences.
    The left-hand panel shows the fit for the signal dataset; right is for the noise dataset.
    The horizontal axis is the log signal evidence while the vertical axis is the log noise evidence.
    }
    \label{fig:GMM} 
\end{figure*}

Our extrapolation uses a Gaussian Mixture Model (GMM)~\cite{Reynolds2009} to fit the distributions of signal and noise evidences [$\pi({\cal Z}_S), \pi({\cal Z}_N)$] using the (1000, 300) data segments that we have already simulated. A GMM models the data as a superposition of independent Gaussian distributions.
The GMM fits are displayed in Figs.~\ref{fig:GMM}a (signal dataset) and Figs.~\ref{fig:GMM}b (noise dataset).
In each panel, the horizontal axis is the log signal evidence while the vertical axis is the log noise evidence.
The color bar indicates the probability density.

\begin{figure}[ht]
\centering
 \includegraphics[width=\columnwidth]{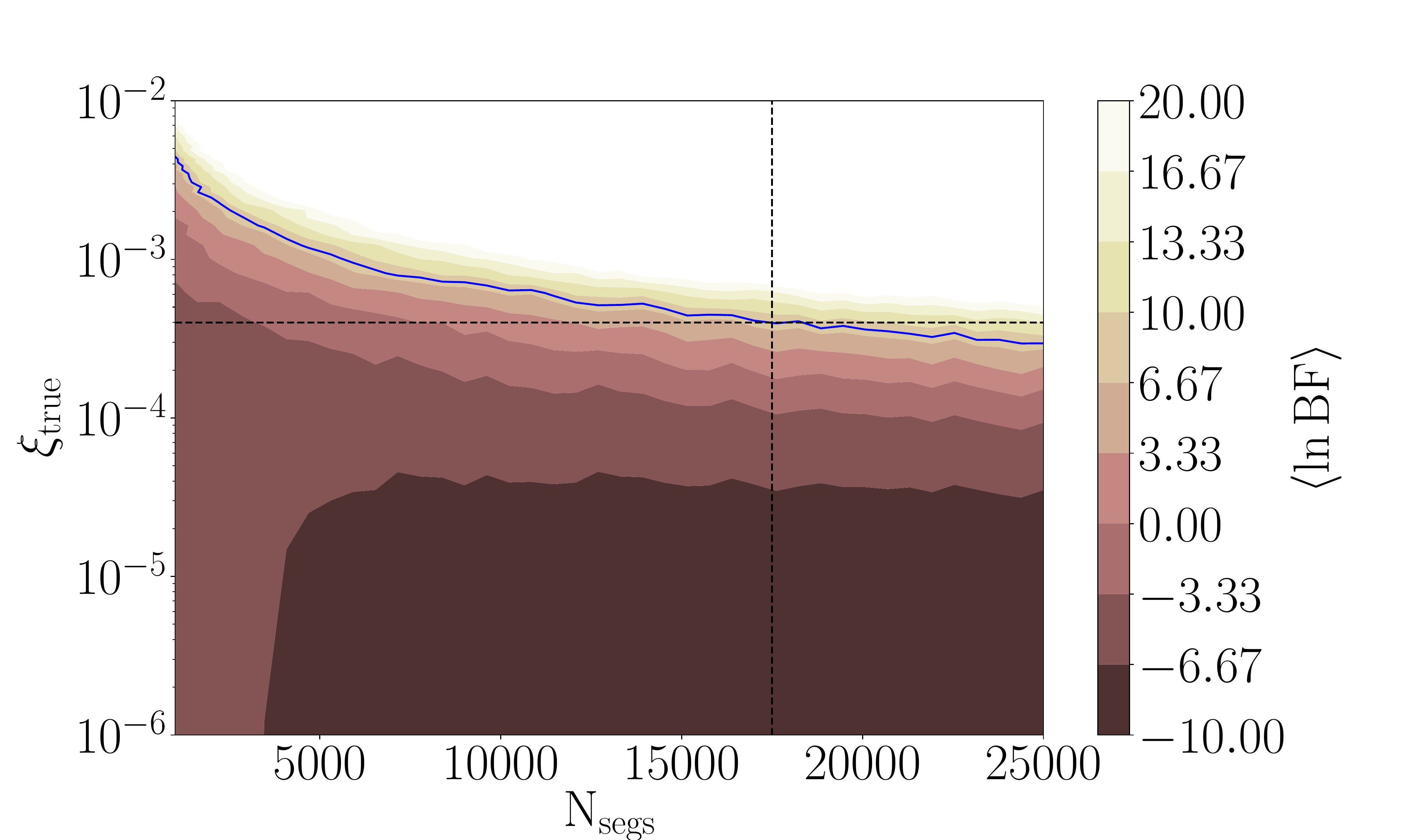}
\caption{
Contours of average $\ln\text{BF}$ as a function of the simulated true duty cycle $\xi_{\text{true}}$ and number of $\unit[4]{s}$ segments $N_{\mathrm{segs}}$.
The blue contour corresponds to $\langle\ln\text{BF}\rangle=8$ where an astrophysical background is detectable.
The color bar saturates at $\ln\text{BF}=20$). The horizontal and vertical lines correspond to $\xi_{\text{true}} = 4 \times 10^{-4}$ and $N_{\mathrm{segs}}=17,500$ respectively.
}
\label{fig:bf_contour}
\end{figure}

Using the GMM fits, we can generate large extrapolated datasets with arbitrarily low duty cycles.
In Fig.~\ref{fig:bf_contour}, we show how the log Bayes factor (Eq.~\ref{eq:bf}) scales with the injected duty cycle $\xi_{\mathrm{true}}$ and the number of segments $N_{\mathrm{segs}}$, the latter of which is equivalent to the observation time.
We average over 1000 realizations of $({\cal Z}_S,{\cal Z}_N)$ drawn from the GMMs (created assuming a two-detector LIGO network operating at design sensitivity).
We find that an astrophysical background with a realistic effective duty cycle $\xi_{\mathrm{true}}\approx\xitrue$ can be detected with  $N_{\mathrm{segs}} \approx \nsegs$ data segments, corresponding to $\unit[20]{hours}$.
This can be compared to a detection time of $\gtrsim\unit[1]{year}$ using cross-correlation~\cite{GW170817_stoch}.
The signal is created by $\approx 7$ sub-threshold events with $z<\zmax$.

We expect that the improvement in sensitivity results from two effects.
First, the likelihood includes information about the non-Gaussian nature of the binary black hole background.
Second, the likelihood includes information about the deterministic nature of compact binary waveforms.
Additional work is required to determine the relative importance of these two factors.

\section{Non-Gaussian noise}\label{noise}
\begin{figure}[t]
\includegraphics[width=\columnwidth]{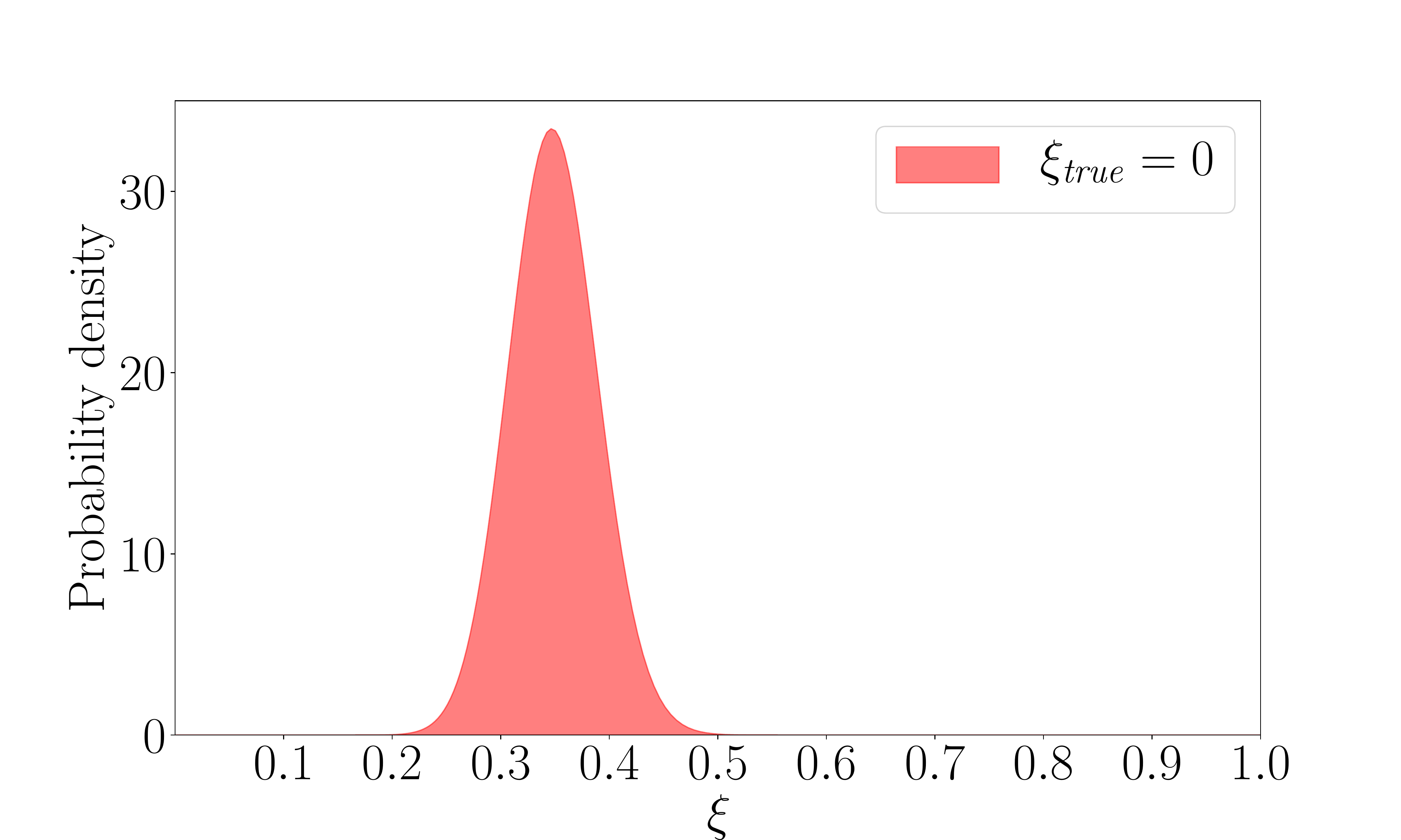}
\caption{
Biased duty cycle posterior, generated with O1 background data and computed using the Gaussian likelihood function, Eq.~\ref{eq:GenLike}.
}
\label{fig:O1_biased}
\end{figure}

Up until this point, we have chosen to model our data as either Gaussian noise or Gaussian noise + signal.
However, real gravitational-wave detectors are subject to non-Gaussian transient noise called glitches.
It is therefore necessary to extend our algorithm to account for non-Gaussian noise in order to ensure that glitches are not mistaken for gravitational-wave signals.
Failure to account for glitches can bias the duty cycle posterior, or worse, yield a false positive. 
We compute the duty cycle posterior for a set of 600 $\unit[4]{s}$ segments of  data from LIGO's first observing run (O1).
We introduce an unphysical time shift to ensure that any real gravitational-wave signals do not produce coherent signals in the Hanford and Livingston detectors.
The true duty cycle should therefore be zero, but the posterior peaks at around $\xi=0.4$ and excludes zero.
Non-stationary noise is producing a significant bias. An example of the undesirable effect of glitches is shown in Fig.~\ref{fig:O1_biased}.

In order to take into account non-Gaussian noise, we extend the likelihood expression from Eq.~\ref{eq:marginalized} to include contributions from glitches. 
For simplicity, we consider a two-detector network, but the results generalize.
For our glitch model, we conservatively suppose that glitches look exactly like binary black hole waveforms except that the waveform in one detector is completely uncorrelated with the waveform in the other.
Any part of a glitch that is orthogonal to the binary black hole signal manifold will not contribute any signal evidence $\mathcal{Z}_S$.

Some additional notation is necessary.
We introduce  parameters $\xi_g^{(1)}$ and $\xi_g^{(2)}$ corresponding to the glitch duty cycle in detectors one and two respectively.
The variables ${\cal Z}^{(1)}_g$ and ${\cal Z}^{(2)}_g$ are the single-detector evidences for the glitch hypothesis given by
\begin{eqnarray}
{\cal Z}^{(1)}_g \equiv & \int d\theta^{(1)} \, 
{\cal L}(s^{(1)}|\theta^{(1)}) \, 
\pi(\theta^{(1)})\\
{\cal Z}^{(2)}_g \equiv & \int d\theta^{(2)} \, 
{\cal L}(s^{(2)}|\theta^{(2)}) \, 
\pi(\theta^{(2)})\,.
\end{eqnarray}
Here, $\theta^{(1)}$ and $\theta^{(2)}$ are the signal parameters for (uncorrelated) glitches in detectors 1 and 2 respectively.
The variables $s^{(1)}$ and $s^{(2)}$ are the strain data in each detector.
We note that the glitch evidences are equivalent to single-detector {\em signal} evidences given our conservative glitch model.
A key feature of this glitch model is that the parameters in each detector are independent.
We introduce the variables ${\cal Z}^{(1)}_N$ and ${\cal Z}^{(2)}_N$ for the single-detector noise evidences (see Eq.~\ref{eq:ZN}).
The variables ${\cal Z}_{S+g}^{(1)}$ and ${\cal Z}_{S+g}^{(2)}$ are the two-detector evidences for a coherent signal with a glitch in one detector.
Finally, the variable ${\cal Z}_{S+g}^{(1,2)}$ is the two-detector evidence for a coherent signal with a glitch in both detectors.
We do not include explicit expressions for ${\cal Z}_{S+g}^{(1)}$, ${\cal Z}_{S+g}^{(2)}$, and ${\cal Z}_{S+g}^{(1,2)}$ because---as we shall see momentarily---they can be approximated as zero.
For the sake of compact notation, we suppress the segment number index $i$ on every evidence term.

\begin{figure*}
\includegraphics[width=2\columnwidth]{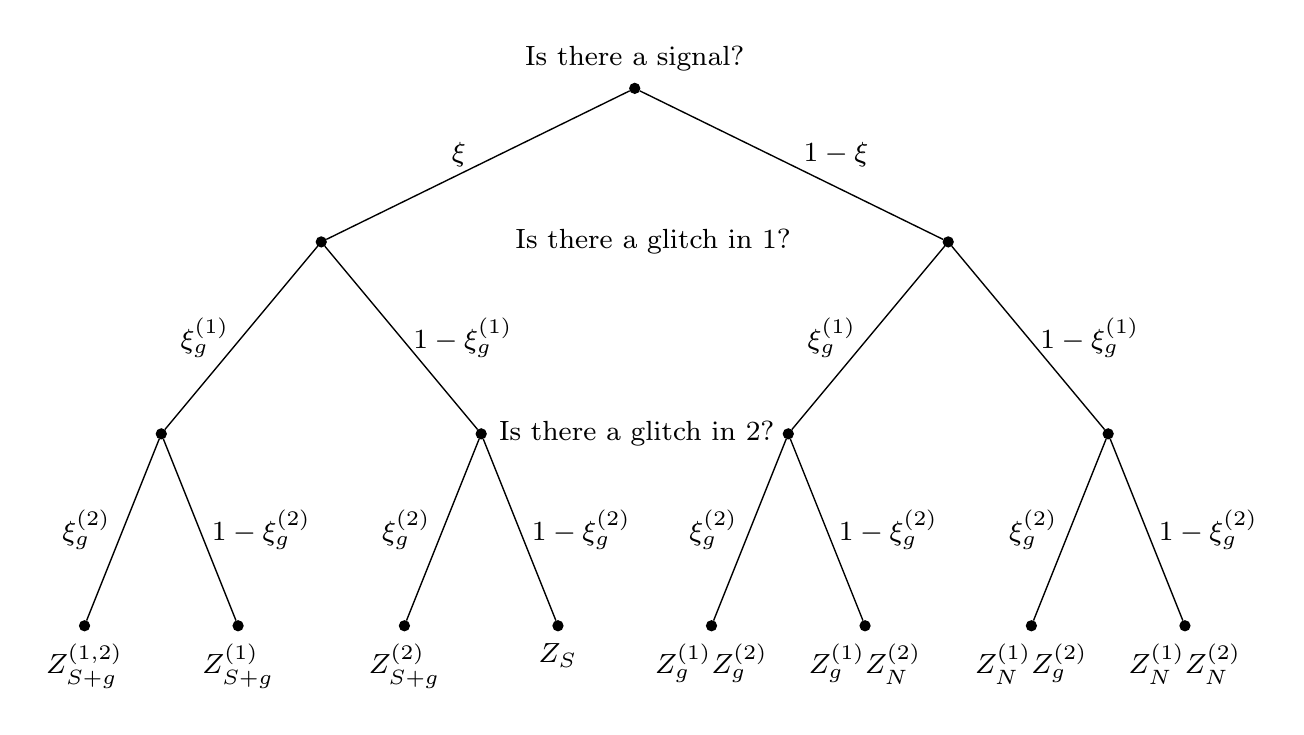}
\caption{
Probability tree for Eq.~\ref{eq:likelihood_full} assuming a two detector network.
The left branches correspond to ``yes'' and the right branches ``no''.
The probability of each branch is labeled accordingly.
The ``leaves'' at the base of the tree show the evidence associated with each path.
}
\label{fig:tree}
\end{figure*}

Given our new definitions, the likelihood for coherent merger events in glitchy data is
\begin{align}
\label{eq:likelihood_full}
\mathfrak{L}(\vec{s}_i|\xi,\xi_g^{(1)},\xi_g^{(2)}) 
= & \xi \left(1-\xi_g^{(1)}\right) \left(1-\xi_g^{(2)}\right) {\cal Z}_S + \nonumber\\
& \left(1-\xi\right)\left(1-\xi_g^{(1)}\right) \left(1-\xi_g^{(2)}\right){\cal Z}_N + \nonumber\\
& \left(1-\xi\right)\,\xi_g^{(1)}\,\left(1-\xi_g^{(2)}\right)\,{\cal Z}_{g}^{(1)} \, {\cal Z}_{N}^{(2)} + \nonumber\\
& \left(1-\xi\right)\,\left(1-\xi_g^{(1)}\right)\,\xi_g^{(2)}\,{\cal Z}_{N}^{(1)} \, {\cal Z}_{g}^{(2)} + \nonumber\\
& \left(1-\xi\right)\,\xi_g^{(1)}\,\xi_g^{(2)}\,{\cal Z}_{g}^{(1)} \, {\cal Z}_{g}^{(2)} + \nonumber\\
& \xi\,\xi_g^{(1)} \left(1-\xi_g^{(2)}\right) {\cal Z}_{S+g}^{(1)} + \nonumber\\
& \xi\,\left(1-\xi_g^{(1)}\right)\xi_g^{(2)} \, {\cal Z}_{S+g}^{(2)} + \nonumber\\
& \xi\,\xi_g^{(1)}\,\xi_g^{(2)} {\cal Z}_{S+g}^{(1,2)}
\end{align}
Each line corresponds to a distinct possibility. A probability tree corresponding to the likelihood function is shown in Fig.~\ref{fig:tree}.
Starting from the first line and reading toward the bottom, the possibilities are: 
\begin{enumerate}
\item signal and no glitches
\item no signal and no glitches
\item no signal and a glitch in detector (1)
\item no signal and a glitch in (2)
\item no signal and a glitch in (1) and (2)
\item signal and a glitch in (1)
\item signal and a glitch in (2)
\item signal with glitches in (1) and (2) .
\end{enumerate}
Note the priors for $\xi, \xi_g^{(1)}, \xi_g^{(2)}$ all run from $(0,1)$.
They are all independent.
Each of the above possibilities corresponds to a branching path in Fig.~\ref{fig:tree}.

For the sake of simplicity, we hypothesize that the last three terms in this likelihood  contain evidences that are small enough to safely ignore in practice: ${\cal Z}_{S+g}^{(1)}$, ${\cal Z}_{S+g}^{(2)}$, and ${\cal Z}_{S+g}^{(1,2)}$.
We expect these three ${\cal Z}$ to be small because they employ overzealous models, which tend to overfit the data.
While terms like ${\cal Z}_S$, ${\cal Z}_g^{(1)}$ and ${\cal Z}_g^{(2)}$ employ 15 parameters to fit a merger event or merger-like glitch, ${\cal Z}_{S+g}^{(1)}$ and ${\cal Z}_{S+g}^{(2)}$ employ 30  parameters to simultaneously fit a merger event {\em and} a glitch.
The ${\cal Z}_{S+g}^{(1,2)}$ term employs 45 parameters to fit a merger event and two glitches.
Since we expect that mergers and glitches are easily fit with 15-parameter models, the final three ${\cal Z}$ incur large Occam factors, which results in small ${\cal Z}$.
This reasoning may not apply to the special case of data that actually contains a merger and a glitch, but we expect such events to be rare.
The recent binary neutron star event GW170817 {\em was} detected coincident with a significant glitch~\cite{GW170817}, but this does not change our expectations.
Binary neutron star signals are in band for $\gtrsim\unit[100]{s}$ versus $\lesssim\unit[0.3]{s}$ for high-mass binary black holes, and so the chance of a coincident glitch is relatively higher.
Of course, one is free to retain the $S+g$ terms, but this requires modification of \lal.
We anticipate that such modifications will be well worth pursuing for a number of applications, including work to extend this analysis to low-mass systems.
The subtraction of a glitch associated with GW170817~\cite{GW170817} using a wavelet reconstruction is a step in this direction~\cite{BayesWave}.

If we assume that the final three ${\cal Z}$ are small enough to ignore, the glitchy likelihood becomes
\begin{align}
\label{eq:Glitchly_likelihood}
\mathfrak{L}(\vec{s}|\xi,\xi_g^{(1)},\xi_g^{(2)}) 
\approx & \xi \left(1-\xi_g^{(1)}\right) \left(1-\xi_g^{(2)}\right) {\cal Z}_S + \nonumber\\
& \left(1-\xi\right)\left(1-\xi_g^{(1)}\right) \left(1-\xi_g^{(2)}\right){\cal Z}_N + \nonumber\\
& \left(1-\xi\right)\,\xi_g^{(1)}\,\left(1-\xi_g^{(2)}\right)\,{\cal Z}_{g}^{(1)} \, {\cal Z}_{N}^{(2)} + \nonumber\\
& \left(1-\xi\right)\,\left(1-\xi_g^{(1)}\right)\,\xi_g^{(2)}\,{\cal Z}_{N}^{(1)} \, {\cal Z}_{g}^{(2)} + \nonumber\\
& \left(1-\xi\right)\,\xi_g^{(1)}\,\xi_g^{(2)}\,{\cal Z}_{g}^{(1)} \, {\cal Z}_{g}^{(2)} .
\end{align}
Below, we show that the ``small-${\cal Z}$'' approximation works well when we apply this likelihood to real data.
This is convenient because it will enable us to obtain results using only existing 15-dimensional signal models.

We now show that our method is robust when applied to real LIGO noise.
Because the LIGO detectors cannot be shielded from gravitational-wave signals, we ensure that the analyzed data cannot contain coincident real signals by performing ``time slides'' in which a relative time offset, longer than the light-travel time between sites, is applied between the data from the detectors.
Time-slides are a common boot-strap technique for generating realistic background noise.

\begin{figure}
        \centering
        \begin{subfigure}[t]{\columnwidth}       \includegraphics[width=\columnwidth]{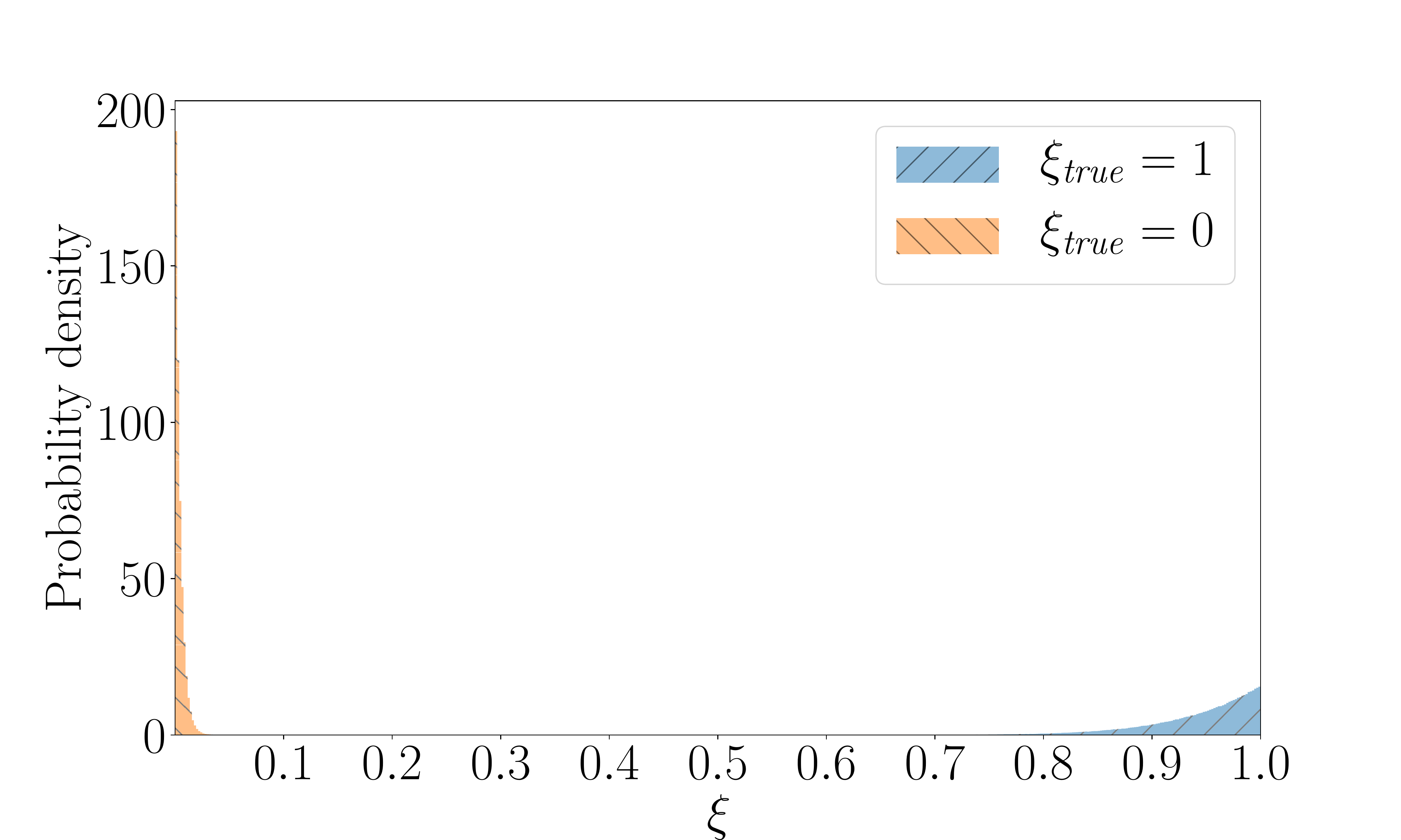}
                \caption{
                Safe and effective duty cycle posteriors for O1 background.
                }
        \end{subfigure}
        \begin{subfigure}[t]{\columnwidth}  \includegraphics[width=\columnwidth]{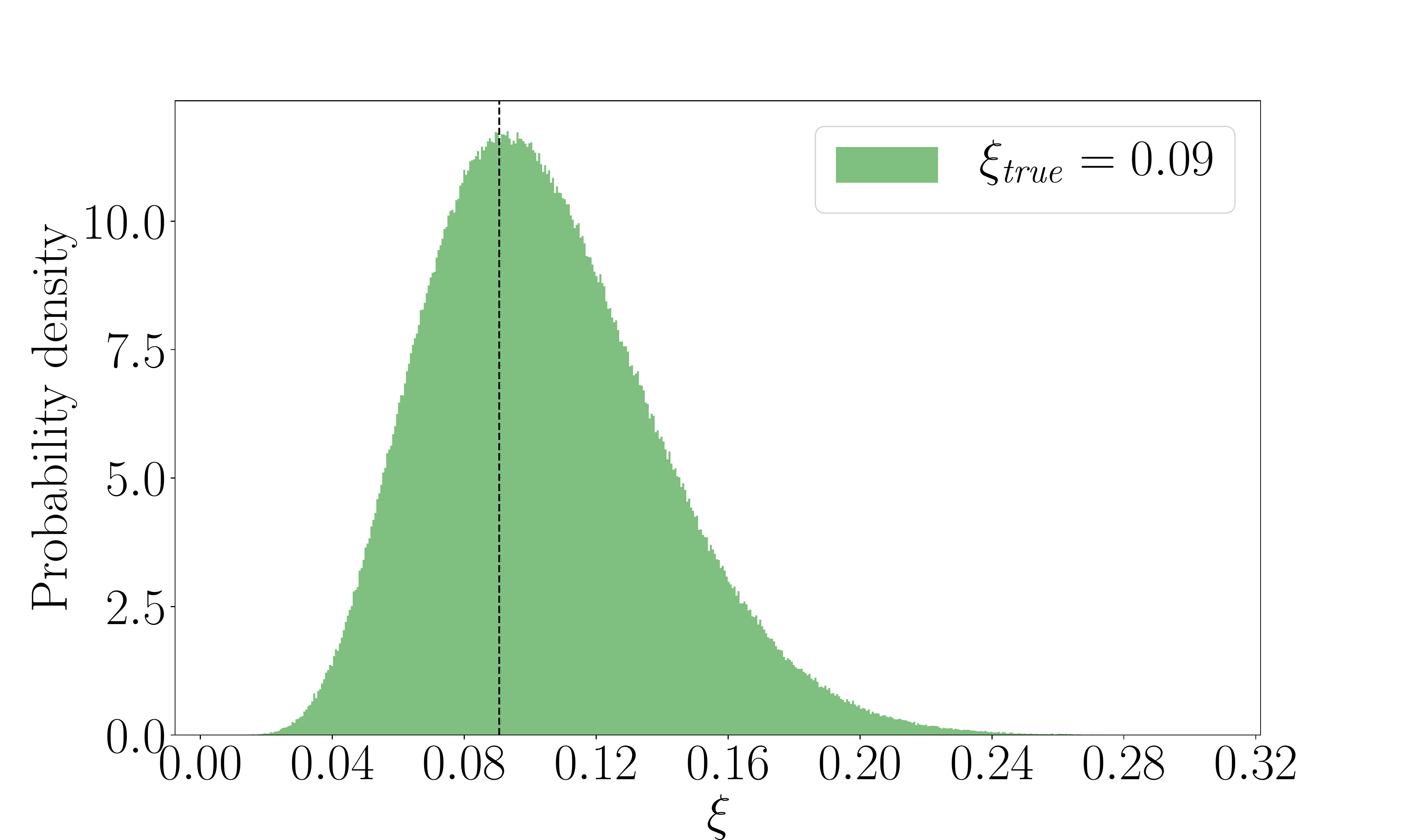}
                \caption{
                Unbiased duty cycle posterior for O1 background containing software injections with a duty cycle $\xi_{\mathrm{true}}=0.09$. The vertical line shows the true duty cycle.
                }
        \end{subfigure}
        \caption{
        Duty cycle posteriors computing using the ``glitchy'' likelihood function Eq.~\ref{eq:Glitchly_likelihood}
        }
\label{fig:O1_effective_safe_unbiased}
\end{figure}

As before, we generate two datasets: a noise-only dataset consisting of 670 $\unit[4]{s}$ background data segments from O1; and an injection dataset consisting of 60 software injections into $\unit[4]{s}$ background data segments from O1.
The signals are generated according to the prescription described in Sec.~\ref{mdc}.
Injections are drawn from a uniform-in-volume distribution.
They are all sub-threshold.
We construct three populations corresponding to $\xi=(0,0.09,1)$ to test for effectiveness, safety, and bias.
We compute the duty cycle posterior using the ``glitchy'' likelihood function (Eq.~\ref{eq:Glitchly_likelihood}). 
In Fig.~\ref{fig:O1_effective_safe_unbiased} we demonstrate that the glitch model employed in the likelihood function Eq.~\ref{eq:Glitchly_likelihood} is safe, effective and unbiased as required.

\section{Computational requirements}\label{computing}
Figure~\ref{fig:bf_contour} implies that, at design sensitivity, a realistic astrophysical background with an effective $z<\zmax$ duty cycle of $(\xi_\text{true}=\xitrue)$ could be detected using $\approx \nsegs$ $\unit[4]{s}$ data segments.
At the time of writing, a typical run-time of \lal\ on $\unit[4]{s}$ data segments is usually no more than $\unit[10]{CPU\,hours}$ \cite{PhysRevD.94.044031}.
To produce evidences in a two detector network, we perform three separate runs: a coherent analysis which produces $(\mathcal{Z}^{i}_S,\mathcal{Z}^{i}_{N})$; and two incoherent analysis which produce $(\mathcal{Z}^{i,(j)}_S,\mathcal{Z}^{i,(j)}_N)$ where the index $j$ labels the detector, $j=(1,2)$.
We therefore estimate that an astrophysical background is detectable in $\unit[500K]{CPU\,hours}$.
This works out to one week with 3000 dedicated CPUs.
Detecting an astrophysical background of binary black holes is therefore feasible with current computing resources of the LIGO Data Grid (LDG), which consists of tens of thousands of CPUs.
This cost estimate assumes relatively high-mass signals split into $\unit[4]{s}$ data segments.
Additional work is required to investigate the cost and performance of the search as the analysis is extended to lower-mass events.
As the minimum mass is decreased, the event rate is expected to increase while the signal duration increases, resulting in a higher duty cycle.
We discuss some of the associated challenges below in Subsection~\ref{BNS}.

While it may be possible to observe a stochastic background by analyzing a small amount of data (with a small computational cost), there is strong motivation for analyzing all available data.
By analyzing an entire observing run, we can make inferences about the population properties of binary black holes (and with additional work, eventually binary neutron stars).
We describe this population inference in the subsequent section.
Carrying out full parameter estimation on a year of data with existing software and hardware would require around 15K continuously operational CPUs in order to process the data in real time.
While this is technically possible, it would be a costly proposition.

In order to reduce the cost, it is worthwhile to consider the development of new computational methods and implementation on new hardware architecture in order to realize the ultimate goals of the search.
These advances might come, for instance, by exploring greater parallelization of parameter estimation methods through the use of, e.g., graphical processor unit (GPU) clusters or larger supercomputer clusters; see, e.g., \cite{BOSS_LDG}.
We note in passing that it might be possible to detect some individually resolvable sources using parameter estimation if they were just below the threshold for detection with matched filter pipelines \cite{BCR}.

\section{Population inference}\label{hyper}
In addition to inferring the the astrophysical duty cycle, which can be related to the local merger rate (Subsection~\ref{rate}) and GW energy density (Subsection~\ref{Omega}), we can also use this framework to infer properties of the binaries that contribute to the astrophysical background.
These population properties may be encoded as hyper-parameters, which affect the prior distributions for various binary parameters.
For example, instead of assuming a flat prior on chirp mass, we can assume that the total mass follows a power-law distribution with some spectral index $\alpha$
\begin{align}
\pi(M) \rightarrow \pi(M|\alpha) 
\propto M^\alpha .
\end{align}
Using Bayesian hierarchical modeling~\cite{Gelman,PhysRevD.86.124032}, we can marginalize over $M$ to obtain a posterior on $\alpha$.

Before sketching how this works, we note that there are three good reasons for eventually developing a sophisticated hyper-parameterization scheme.
First, there is interesting information encoded in the population properties of binary black hole coalescences, which can be used, for example, to study binary black hole formation channels; see,~e.g.,~\cite{spin,Stevenson,Mandel,Maya,Vitale,Stevenson2}.
Second, we should hyper-parameterize theoretically uncertain prior distributions in order to obtain an unbiased estimate of the coalescence rate.
For example, a systematic error in the assumed mass spectrum will lead to a systematic error in the inferred rate posterior.
Finally, by acknowledging theoretical uncertainty with hyper-parameterization, we should improve the sensitivity of the search. This should generally be true when \textit{necessary} parameters are added to the search.
For example, marginalizing over the unknown mass spectrum index is likely to yield a higher Bayes factor than just assuming some value (unless we have a strong prior belief in some particular value). 

The likelihood function for a vector of hyper-parameters $\Lambda$ is obtained by introducing a conditional prior $\pi(\theta|\Lambda)$ and then marginalizing over $\theta$
\begin{eqnarray}
{\cal L}(h|\Lambda) &=& \int d\theta {\cal L}(h|\theta) \pi(\theta|\Lambda)
\end{eqnarray}
The distribution $\pi(\theta|\Lambda)$ is the hyper-parameterized conditional prior. The above integral can be computed without performing any extra sampling by ``recycling'' the posterior samples already computed by \lal. 
Replacing the likelihood by a sum over samples we obtain:
\begin{align}\label{eq:hyper}
{\cal L}(h|\Lambda) \approx \prod_i^n
\frac{1}{n_i} \sum_{k=1}^{n_i}
\frac{\pi(\theta_{i,k}|\Lambda)}{\pi(\theta_{i,k}|\text{\sc LAL})} .
\end{align}
The distribution $\pi(\theta|\text{\sc LAL})$ is the prior distribution used for the generation of the posterior samples by \lal.
The product over $i$ runs over the number of data segments from $1$ to $n$.
The sum over $k$ runs over the number of posterior samples from $1$ to $n_i$.
(Each segment has a different number of posterior samples.)

We note that Eq.~\ref{eq:hyper} can be applied as a post-processing step.
That is, one need only run \lal\ once.
As long as we use relatively uniformative priors for $\pi(\theta|\text{\sc LAL})$, the resulting posterior samples can be reweighted in order to obtain the likelihood for different hyper-parameter values.
The posterior for $\Lambda$ and the (hyper-marginalized) Bayes factor are calculated the usual way.
The first prior distributions to hyper-parameterize are (1) those, which shed light on the mechanisms of binary black hole formation and (2) those, which are subject to significant theoretical uncertainty.
Probably, this means hyper-parameter descriptions of the binary black hole mass spectrum and the distribution of black hole spins.
The former is the subject of work in preparation~\cite{CallisterPop}.

\section{Extensions of the search}
\label{extensions}
In this section, we consider five possible extensions to this search method: searches for binary neutron stars, continuous waves, super-massive black hole binaries, bursts, and glitch classification.
We also consider the simultaneous estimation of Gaussian background with a non-Gaussian foreground, but the discussion requires some depth, and so we discuss that separately in the next section.
Our goal here is two-fold.
First, we seek to illustrate promising directions for future research.
Second, we endeavor to showcase that, for every astrophysical background, there is an appropriately complete likelihood function, which can serve as the basis for an optimal search.

\subsection{Binary neutron stars.}\label{BNS}
An obvious extension of this technique is to search for backgrounds of binary neutron stars.
Recent observations of a binary neutron star inspiral~\cite{GW170817} suggest that the gravitational-wave energy density $\Omega_\text{gw}$ from such systems is roughly comparable to the energy density from binary black holes~\cite{GW150914_stoch}.
Binary neutron stars are less massive than typical binary black holes, but they coalesce more frequently.
Thus, while the two backgrounds are expected to produce comparable $\Omega_\text{gw}$, the rate of binary neutron star mergers is much higher (by a factor of $\approx17$).
Moreover, binary neutron star waveforns are much longer than binary black hole waveforms.
The first binary black hole event GW150914 was in-band for $\approx\unit[0.3]{s}$~\cite{GW150914} versus $\unit[100]{s}$ for the first binary neutron star GW170817~\cite{GW170817}.
It is important to understand that while the binary neutron star background is {\em continuous} (always present), it is still {non-Gaussian}: the mergers are, for the most part, clearly separated in time~\cite{Rosado}.

These two effects all but ensure that a signal is present in the data at any given time.
At any given moment, there are likely to be 15 binary neutron stars somewhere in the Universe, producing gravitational waves with $f>\unit[10]{Hz}$ (versus 0.06 for binary black holes)~\cite{GW170817_stoch}.
This violates an assumption in our formulation of the binary black hole search: that the vast majority of segments contain either no signal or one signal.

All is not lost.
There are probably a number of solutions.
One possibility is to treat the number of events in some data segment as a free parameter using the local rate to inform the prior.
The formalism of reversible-jump Markov chain Monte Carlo, where the dimensionality of the parameter space is itself a free parameter, is potentially suited for this problem; see, e.g.,~\cite{BayesWave}.
Using this plan of attack, the algorithm could attempt, for example, to simultaneously fit dozens of binary neutron signals simultaneously, each with 15 parameters.
It is not clear if such a high-dimensional search would be computationally feasible, but further investigation is warranted.

\subsection{Continuous waves}
There are thought to be $\approx50000$ isolated neutron stars in the Milky Way emitting gravitational waves in the observing band of advanced detectors~\cite{Lorimer,ns_sgwb}.
The computational power required to search the full parameter space is so vast that semi-coherent techniques are required.
The data are analyzed coherently in small chunks of manageable size, typically $\unit[1800]{s}$~\cite{PowerFlux}.
The results from each chunk are combined incoherently.

Since the signals overlap in time, it would not make sense to define a duty cycle as fraction of segment that include a signal.
However, it may be possible to define a duty cycle equal to the fraction of {\em frequency bins} that contain a signal.
To get a (very rough) idea of the duty cycle, we can assume that continuous-wave sources are roughly evenly distributed throughout the advanced detector observing band from $\unit[10-1500]{Hz}$~\cite{PowerFlux}.
Assuming $\unit[1800]{s}$ segments (with frequency resolution $\unit[0.56]{mHz}$), there are 2.7M frequency bins.
This implies a duty cycle of $\xi\approx 2\%$, which is much less than one.
We therefore expect that this formalism can be applied relatively straightforwardly to search for a population of Galactic neutron stars.
This proposal bears similarities to~\cite{Fan}, which proposes an ensemble search for {\em known} pulsars.

\subsection{Super-massive black hole binaries}
The background from super-massive black hole binaries is analogous to the background from isolated neutron stars, except the measurements are carried out by pulsar timing arrays; for a review see~\cite{Hobbs,RosadoSMBH}.
Most super-massive black hole binaries are expected to produce nearly monochromatic signals that evolve slowly over the decade-long observation period.
At any one time, there might be $10^4-10^5$ such binaries emitting gravitational waves in the pulsar-timing band~\cite{sesana}.
The probability that an inspiralling binary emits gravitational waves on the interval $(f, f+df)$ is \footnote{The probability distribution $\pi(f)$ can be derived from the post-Newtonian approximation for the time-to-coalescence from a frequency $f$, sometimes known as the ``chirp time''. To leading order, the chirp-time $\tau$ is related to frequency by $\tau \propto f^{-8/3}$ \cite{Blanchet2014}. By conservation of probability $\pi(f) df = \pi(\tau(f))d\tau$, and assuming $\pi(\tau)$ is constant, which is to say there is no preferred coalescence time, we arrive at  $\pi(f) =f^{-11/3}$.} 
\begin{align}
\pi(f) \propto f^{-11/3} .
\end{align}
Thus, most binaries emit near the lower limit of the observing band.
Pulsar timing arrays observe in a band $\unit[1-100]{nHz}$ with a resolution of $\approx\unit[1]{nHz}$.
We expect 90\% of the binaries to be observed in a narrow band of $(\unit[1-2.4]{nHz})$.
This is also the most sensitive part of the band.
Thus, we expect many thousands of binaries per frequency bin in the relevant part of the band.

We are therefore unable to define a duty cycle as the fraction of frequency bins containing a signal.
The frequency bins that contribute the signal do not contain {\em a} signal, they contain thousands.
One can imagine defining the duty cycle in terms of sky location: the fraction of patches of sky which contain a signal.
However, given current pulsar timing arrays, it seems unlikely that there are enough quasi-independent patches of sky so that the expected duty cycle is less than one; see, e.g.,~\cite{Zhu}.
If so, and if we are not missing some other means of distinguishing supermassive black hole binary signals, it seems that the pulsar timing background is, for all intents and purposes, Gaussian in nature, at least as measured by foreseeable detectors.

\subsection{Bursts}
Gravitational-wave bursts are unmodeled transients, which can be contrasted with well-modeled signals from compact binaries.
There are expected bursts from objects like supernovae,  but gravitational-wave astronomers also search for unexpected bursts.
Given the significant theoretical uncertainties about the loudness and rate of different gravitational-wave bursts, it is hard to say if there is a significant stochastic background from bursts, and if so, whether or not it is Gaussian.
If, for example, there is a detectable background from supernovae~\cite{Crocker}, one might expect a Gaussian background since supernovae explode in the Universe at a rate of $\approx\unit[30]{s^{-1}}$.
Alternatively, the burst background might be dominated by louder but less frequent signals creating a non-Gaussian background from, e.g., cusps of cosmic strings.

Given these theoretical uncertainties, it seems worthwhile to carry out a non-Gaussian search for bursts.
The formalism described here can be  extended.
Instead of marginalizing over compact binary parameters, one can imagine marginalizing over arbitrary combinations of wavelets, see, e.g.,~\cite{BayesWave}.
Burst waveforms that can be easily parameterized, such as cosmic string bursts~\cite{Damour} and gravitational-wave memory~\cite{Favata}, are relatively straightforward to implement in this formalism.

\subsection{Glitch classification}
The method can also be repurposed to identify populations of glitches.
To do this, we employ the formalism for non-Gaussian noise.
We assume that the glitches are described by a hyper-parameterized prior.
Following the method described in Section~\ref{hyper}, we can estimate the hyper-parameters of the glitch population in order to classify populations of glitches.
For example, one might find that there is a population of transients in the LIGO Hanford detector that are best fit by templates corresponding to $50+50 \, M_\odot$ mergers.
This idea is only a sketch, but we can envision developing a practical tool, which could be useful for commissioning and detector characterization.
For a related discussion, see~\cite{powell}.

\section{Simultaneous estimation of  Gaussian background with a non-Gaussian foreground}\label{Gaussian}
In the long run, we are interested in uncovering the primordial background likely to be lurking underneath the astrophysical background from compact binary mergers.
Measurement of a primordial background is considered a Holy Grail of gravitational-wave astronomy, potentially allowing us to probe times well before the formation of the cosmic microwave background and to test energy scales that are not accessible through any other means; see, e.g.,~\cite{Maggiore}.
The problem of disentangling primordial backgrounds from astrophysical foregrounds is therefore important.
The primordial background is likely to be Gaussian.
In this section, we sketch out how the analysis could be extended to simultaneously measure a Gaussian background in the presence of a non-Gaussian foreground.
For the sake of readability, we suppress frequency dependence as well as indices denoting segment number.
The reader should consider both of these to be implied.

As a first step, we derive the likelihood for a purely Gaussian background.
This derivation will be helpful in order to see how the result is generalized to include simultaneous Gaussian and non-Gaussian signals.
The likelihood of obtaining strain data $s$ given a persistent Gaussian signal ${\bf h_G}$ is
\begin{align}\label{eq:stoch}
\log\big[{\cal L}({\bf s} | {\bf h_G})\big] \propto
-\frac{1}{2} \langle {\bf s} - {\bf h_G}, {\bf s}-{\bf h_G}\rangle .
\end{align}
The inner product, defined in Eq.~\ref{eq:inner}, includes an implicit sum over detectors.
Variables in bold-face are vectors with a different entry for each detector, e.g., ${\bf h_G}=(h_G^{(1)}, h_G^{(2)}, ...)$.
The strain induced in two detectors $\alpha$ and $\beta$ is not in general the same, though, the two strains are related via the overlap reduction function $\gamma_{\alpha \beta}$~\cite{christensen}:
\begin{align}\label{eq:gamma}
\langle (h_G^{(\alpha)})^\dagger h_G^{(\beta)} \rangle = \gamma_{\alpha\beta} S_h .
\end{align}

We do not have a template for ${\bf h_G}$ because it is described by a stochastic process.
Our prior for ${\bf h_G}$ is
\begin{align}\label{eq:stoch}
\pi({\bf h_G}|S_h) = & \frac{1}{\det(2\pi{\bf \Sigma})^{1/2}}\exp\left(-\frac{1}{2}{\bf h_G}^\dagger {\bf \Sigma}^{-1} {\bf h_G}\right) \\
{\bf h} = & \left( \begin{array}{c} h_G^{(1)} \\ h_G^{(2)} \end{array} \right) \\
{\bf \Sigma} = & S_h \left( \begin{array}{cc} \gamma_{11} & \gamma_{12} \\ \gamma_{21} & \gamma_{22} \end{array} \right) .
\end{align}
where $S_h$ is the signal power-spectral density.
Note that both $S_h$ and ${\bf h_G}$ are implicit functions of frequency.
(This frequency dependence may be described by additional parameters such as a spectral index.)
The prior in Eq.~\ref{eq:stoch} states that ${\bf h_G}$ is a Gaussian field characterized by a power spectrum $S_h$ and with covariance described by Eq.~\ref{eq:gamma}.

Before proceeding further, we introduce a simplifying assumption: that the detector network consists of two co-located detectors.
This assumption is by no means necessary, but it will facilitate straightforward comparison with~\cite{RomanoCornish}.
Employing this assumption, we can make the following substitutions
\begin{align}
{\bf h_G} \rightarrow & h_G \left(\begin{array}{c} 1 \\ 1 \end{array} \right) \\
{\bf \Sigma} \rightarrow & \gamma S_h \left( \begin{array}{cc} 1 & 1 \\ 1 & 1 \end{array} \right).
\end{align}
Here, $\gamma$ with no indices is the overlap reduction function for a co-located detector pair.
For comparison with~\cite{RomanoCornish}, we work for the time being with ``effective power'' $S_h^\text{eff} \equiv \gamma S_h$.

Next, following~\cite{RomanoCornish} (see their Section~4.2), we marginalize over $h_G$ in Eq.~\ref{eq:stoch} to obtain a marginalized likelihood
\begin{align}\label{eq:CornishRomano}
{\cal L}({\bf s}|S_h^\text{eff}) = \frac{1}{\sqrt{2\pi \det\left({\bf C}\right)}} \exp\left(-\frac{1}{2}{\bf s}^T {\bf C}^{-1} {\bf s}\right) ,
\end{align}
where
\begin{align}
{\bf s} = & \left( \begin{array}{c} s_{1} \\ s_{2} \end{array} \right) \\
{\bf C} = & \left( \begin{array}{cc} P + S_h^\text{eff} & S_h^\text{eff} \\ S_h^\text{eff} & P + S_h^\text{eff} \end{array} \right) .
\end{align}
Following~\cite{RomanoCornish}, we define $P\equiv\sigma^2$.
Remember that there is an implied sum over frequency bins in the expression ${\bf s}^T {\bf C}^{-1} \bf {s}$ and that $\det({\bf C})$ includes an implied product over frequency bins.
After combining data from multiple segments, there is, additionally, an implied sum over segments in the expression ${\bf s}^T {\bf C}^{-1} \bf {s}$ and an implicit product over segments in the expression $\det({\bf C})$.

There are two terms in the exponential of Eq.~\ref{eq:CornishRomano}: an auto-power term containing $s_1^2+s_2^2$ and a cross-power term containing $s_1^* s_2$.
The auto-power terms are considered unreliable because a detection relying on auto-power would require a a precise noise budget.
If the noise power spectral density includes a component that exceeds the noise budget, experimentalists assume it is an unmodeled noise, not a stochastic background.
The typical solution is to start over with a likelihood constructed only out of cross-power~\cite{AllenRomano}
\begin{align}\label{eq:CC}
{\cal L}({\bf s}|S_h^\text{eff}) = \frac{1}{\sqrt{2\pi P}}
\exp\left(-(s_1^*s_2-S_h^\text{eff})^2/2P^2\right) .
\end{align}
This prescription ensures that auto-power from unknown noise does not create a false signal.

\begin{figure}[h]
\includegraphics[width=\columnwidth]{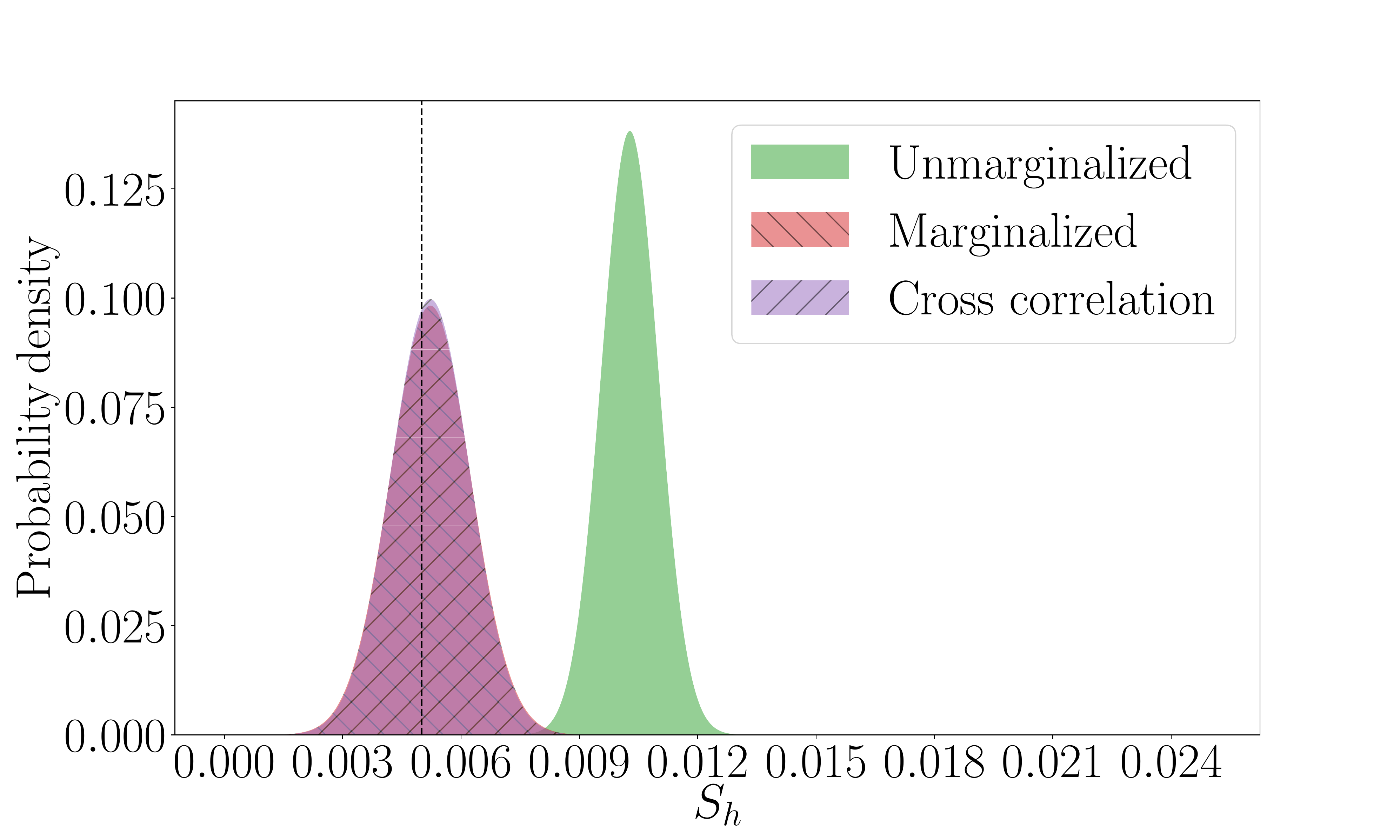}
\caption{
A comparison of likelihood and marginalized likelihood functions for stochastic power spectral density $S_h$.
We carry out a numerical experiment with toy-model data.
We plot likelihood functions ${\cal L}(s|S_h)$ using cross- and auto-power (Eq.~\ref{eq:CornishRomano}, ``Unmarginalized'', green), using cross-power only (Eq.~\ref{eq:CC}, ``Cross correlation'', hatched purple), and using cross- and auto-power, but marginalizing over large uncertainty in the auto-power (Eq.~\ref{eq:mtest}, ``Marginalized'', hatched red).
The purple is nearly indistinguishable from the red.
The green ``unmarginalized'' distribution is narrower, but it is also biased because it does not take into account uncertainty in $P$.
(If we do not include a systematic error in $P$, the posterior peaks in the correct place.)
The horizontal axis is strain power in arbitrary units and the vertical axis is the likelihood.
The black vertical line indicates the injected value.
}
\label{fig:mtest}
\end{figure}

The prescription in Eq.~\ref{eq:CC} does not lend itself to our present purposes.
Fortunately, there is a Bayesian approach that does.
The notion that interferometer noise budgets are not trustworthy enough to detect excess auto-power can be framed in terms of a prior belief.
We treat $P$ as a parameter with a prior distribution $\pi(P)$ peaked at $P_0$ with some width $\sigma_P$, which is wide compared to $S_h$
\begin{align}
\pi(P) \propto \exp(-(P-P_0)^2/2\sigma_p^2) .
\end{align}
That is to say, we do not trust our measurement of the noise at a level comparable to the size of the stochastic signal power $S_h$. 
Marginalizing over $P$, we obtain an evidence (marginalized likelihood) for each $S_h^\text{eff}$, which cannot be tricked by excess auto-power
\begin{align}\label{eq:mtest}
{\cal Z}(S_h^\text{eff}) = \int dP \, \pi(P) \frac{1}{\sqrt{2\pi \det\left({\bf C}\right)}} \exp\left(-\frac{1}{2}{\bf s}^T {\bf C}^{-1} {\bf s}\right) .
\end{align}
When $\sigma_P \gg S_h^\text{eff}$, the stochastic signal encoded in the auto-power is lost, and we recover something close to the cross-correlation likelihood; see Fig.~\ref{fig:mtest}.

In Fig.~\ref{fig:mtest}, we show likelihoods for a demonstration calculation with mock data.
The cross-correlation likelihood (Eq.~\ref{eq:CC}, /-purple) and the marginalized likelihood from Eq.~\ref{eq:mtest} ($\setminus$-red) are nearly identical.
Both are consistent with the injected value of $S_h$.
If we calculate the likelihood from Eq.~\ref{eq:CornishRomano}, but do not marginalize over uncertainty in $P$, we obtain the green distribution.
Since this distribution includes information from both cross- and auto-power, it is narrower than the other distributions.
However, it is also biased because an error in $P$ leads to an overestimate of $S_h$.
If we do not include a systematic error in $P$, the posterior peaks in the correct place.
For the purposes of this sketch, we  model uncertainty in the noise power spectral density with a simple expression for $\pi(P)$.
We note in passing that significantly more sophisticated models are available in order to take into account, e.g., frequency-dependent artifacts such as instrumental lines~\cite{PhysRevD.96.102001}.

Now we are ready to sketch out a method to detect a background with both Gaussian and non-Gaussian components.
The likelihood includes a Gaussian component $h_G$ and a non-Gaussian component $h_\text{NG}(\theta)$
\begin{align}\label{eq:stoch2}
\log\big[{\cal L}({\bf s} | {\bf h_G}, {\bf h_{NG}})\big] \propto -\frac{1}{2} \langle {\bf s} - {\bf h_{NG}} - {\bf h_{G}}, {\bf s}-{\bf h_{NG}} - {\bf h_{G}}\rangle
\end{align}
The non-Gaussian signal depends implicitly on binary parameters $\theta$.
The bold face indicates a vector with entries for different detectors.

Following the same reasoning we used to calculate the Gaussian likelihood in Eq.~\ref{eq:CornishRomano}, the Gaussian + non-Gaussian likelihood is
\begin{align}
{\cal L}({\bf s}|{\bf h_{NG}},S_h) = & \frac{1}{\sqrt{\det 2\pi \left({\bf C}\right)}} \nonumber\\
& \exp\left(-\frac{1}{2}({\bf s}-{\bf h_{NG}})^T {\bf C}^{-1} ({\bf s}-{\bf h_{NG}})\right) .
\end{align}
By introducing a suitable prior $\pi(P)$ and marginalizing over uncertainty in the detector noise, it should be possible to simultaneously infer the existence of a population of unresolved binaries and a continuous Gaussian background.
\begin{align}\label{eq:KitchenSink}
{\cal Z}({\bf h_{NG}},S_h) = & 
\int dP \, \pi(P) {\cal L}({\bf s}|{\bf h_{NG}},S_h) .
\end{align}
We note that, by marginalizing over uncertainty in $P$, we should not hamper our ability to measure binary signals.
Individual binary signals contribute to our evidence by matching with waveform templates, not by inducing excess power.
Indeed, parameterized noise models are a fixture of recent transient analysis~\cite{BayesWave}.

Using the Gaussian + non-Gaussian likelihood function in Eq.~\ref{eq:KitchenSink}, we can marginalize over the implicit binary parameters $\theta$, to obtain signal and ``noise'' evidences analogous to Eqs.~\ref{eq:ZS} and~\ref{eq:ZN}
\begin{align}
{\cal Z}_S(S_h) = &  \int d\theta \, {\cal Z}({\bf h_{NG}},S_h) \\
{\cal Z}_N(S_h) = &  {\cal Z}({\bf h_{NG}=0},S_h) .
\end{align}
These evidences are functions of the hyper-parameter $S_h$, which describes the Gaussian background.
We refer to the ``noise'' evidence with quotation marks because it contains a Gaussian background signal.
In order to calculate them in \lal, one must implement the revised likelihood function (Eq.~\ref{eq:KitchenSink}) with a new parameter $S_h$ and a suitable prior $\pi(P)$.
The next step is to introduce the duty cycle as in Eq.~\ref{eq:GenLike} as in Section~\ref{formalism}.
Retracing our steps, we obtain a posterior $p(\xi,S_h|\{\vec{s}\})$ analogous to Eq.~\ref{eq:p_xi}.
We may convert $S_h$ into $\Omega_{G}$, the Gaussian energy density; see, e.g.,~\cite{locus}
\begin{align}
\Omega_{G}(f) = \frac{2\pi^2 f^3}{3 H_0^2} S_h ,
\end{align}
where $H_0$ is the Hubble parameter.

While significant work is required to go beyond this sketch and demonstrate this technique, we hope this proposal provides a useful outline to construct the optimal method to simultaneously detect Gaussian and non-Gaussian backgrounds.
In particular, we expect the optimal method to improve upon various schemes in which astrophysical events are fit separately and then subtracted, see, e.g.,~\cite{CutlerHarms}.
It is worth investigating as a promising tool for future efforts to measure primordial backgrounds.

\section{Conclusions}\label{conclusions}
Preliminary estimates suggest that advanced detectors, operating at design sensitivity, can detect a stochastic background from binary black holes in about one day.
These estimates rely on extrapolation using Gaussian mixture modeling of our Bayesian evidence distributions.
The next step is to carry out a mock data challenge in which we demonstrate the safety and efficacy of the search using $\approx\unit[1]{day}$ of design sensitivity Monte Carlo data.
Such a demonstration would allow us to verify the extrapolations made here with a modest computational cost $\approx\unit[500\text{K}]{core\,hours}$.

We have highlighted new directions worthy of deeper investigation beyond the overview we provide here.
It will be interesting to more fully develop this method for other audio-band sources of gravitational waves including binary neutron stars, continuous waves,  unmodeled bursts, and glitches.
The method does not appear to be helpful for pulsar timing.
We look forward to demonstrating the simultaneous detection of a Gaussian background in the presence of a non-Gaussian foreground.

This formulation of binary black hole detection provides a unified framework for the analysis of both resolvable signals and a stochastic background of unresolvable signals.
It is also a natural framework to carry out analysis of the population properties of binary black holes.
Since the resolved and unresolved binaries are analyzed as a single dataset, it is possible to eliminate selection effects.

\begin{acknowledgements}
We thank Joe Romano, Letizia Sammut, Tom Callister, Max Isi, Xingjiang Zhu, Pablo Rosado and Kent Blackburn for helpful comments.
RS \& ET are supported by CE170100004.
ET is supported through ARC FT150100281.
LIGO was constructed by the California Institute of
Technology and Massachusetts Institute of Technology
with funding from the National Science Foundation and
operates under cooperative agreement PHY-0757058.
This manuscript has LIGO Document ID LIGO-P1700407.
\end{acknowledgements}

\begin{appendix}
\section{Derivation of $\xi_\text{true}$ for Monte Carlo Study}\label{xi_derivation}
In this section, we justify the choice of $\xi_\text{true}=\xitrue$ as a reasonable choice of duty cycle for our Monte Carlo study in order to simulate a plausible stochastic background.
We show $\xi_\text{true}=\xitrue$ implies an energy density spectrum that is roughly consistent with the prediction of~\cite{GW170817_stoch}, and that such a background cannot be detected with cross correlation in one year of design sensitivity Hanford-Livingston data.

\begin{figure*}[htb] 
  \begin{subfigure}[c]{0.49\linewidth}
    \includegraphics[width=1\linewidth]{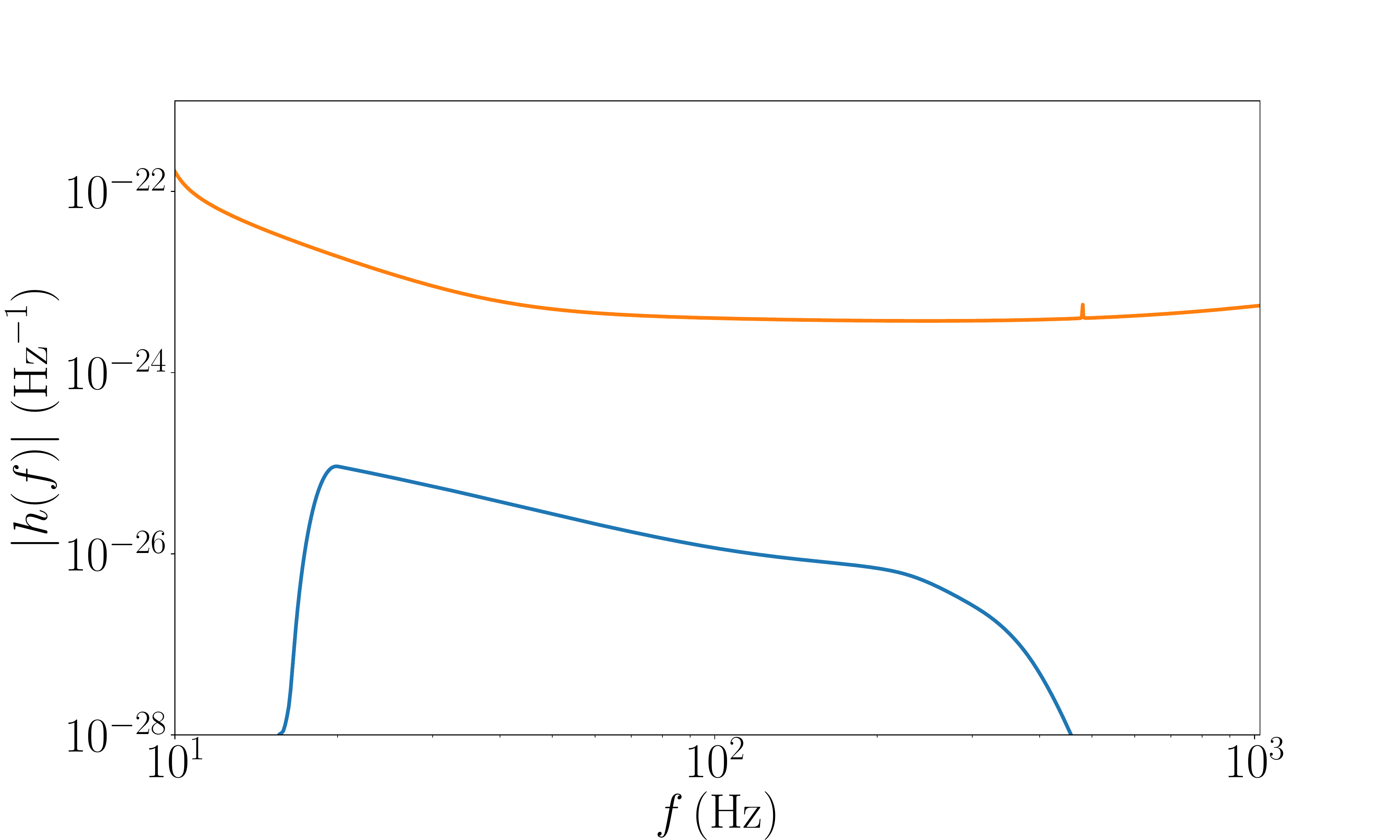} 
  \end{subfigure}
    \begin{subfigure}[c]{0.49\linewidth}
    \includegraphics[width=1\linewidth]{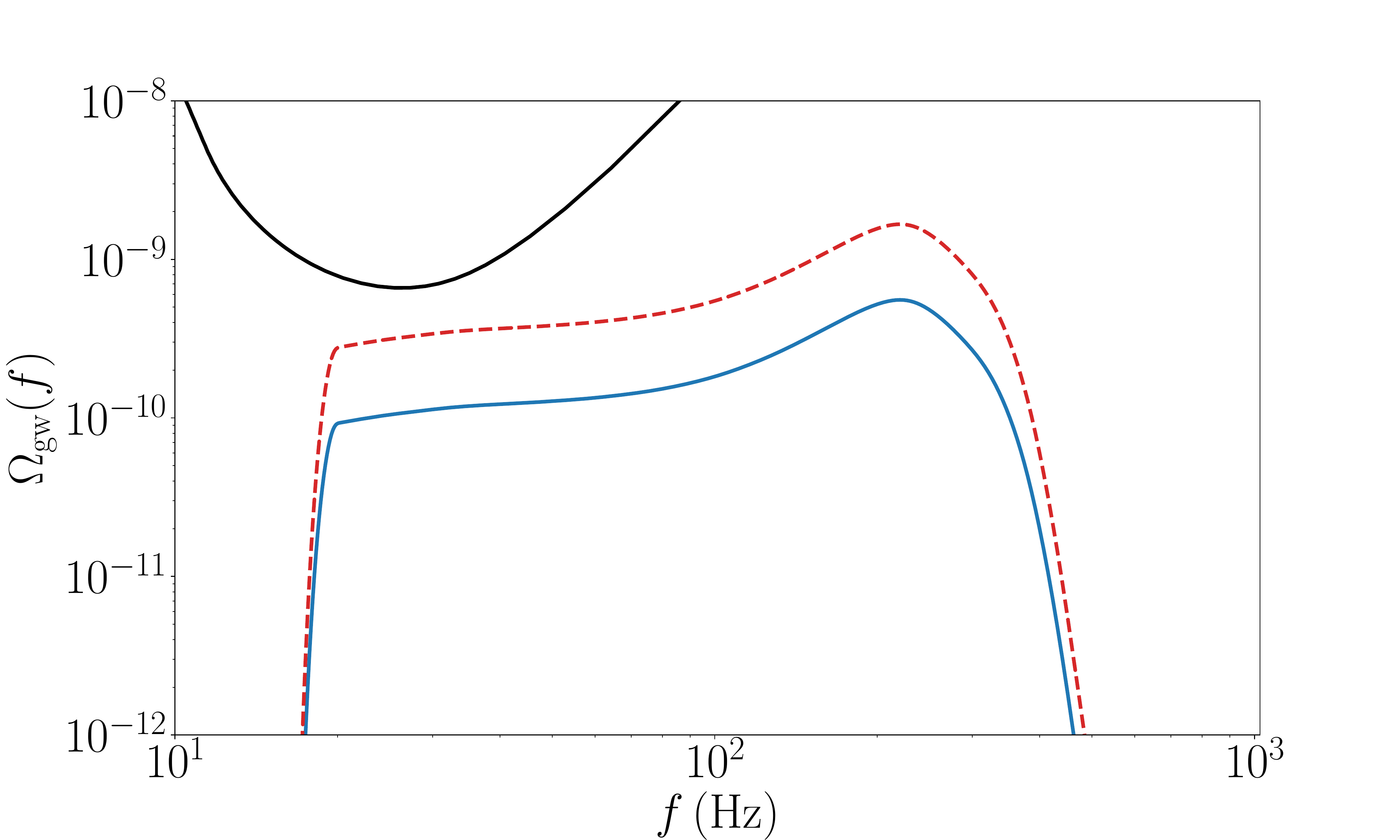} 
  \end{subfigure}
      \caption{
  Left: amplitude spectral density $[S_h(f)]^{1/2}$.
  Blue is for the dataset used in our mock dataset ($z<\zmax$).
  The drop at $\unit[20]{Hz}$ is an artifact from our cut-off frequency.
  The orange curve is the LIGO design sensitivity noise curve.
  Right: energy density $\Omega_\text{gw}(f)$.
   Solid blue is for the dataset used in our mock dataset ($z<\zmax$) while dashed red is the same as solid blue, but including binaries out to arbitrarily high redshifts.
  The dashed red background is three times higher than the solid blue.
  The black curve is the $\unit[1]{\sigma}$ power-law sensitivity curve~\cite{locus}, indicating the sensitivity of the stochastic cross-correlation search assuming a LIGO Hanford-Livingston network operating at design sensitivity for one year.
  Since the red dashed curve is below the black sensitivity curve, the background is not detectable with cross correlation after one year.
    }
    \label{fig:Omega} 
\end{figure*}

The first step is to calculate the average gravitational-wave power-spectral density $S_h(f)$ for data segments in which a signal is present:
\begin{align}
S_h(f) = \int d\theta \, \pi(\theta) \,
\frac{|\tilde{h}_+(f|\theta)|^2 + |\tilde{h}_\times (f|\theta)|^2}{\cal N} .
\end{align}
Here, $\tilde{h}_+(f|\theta)$ and $\tilde{h}_\times(f|\theta)$ are the $+$ and $\times$ components of the Fourier transform of the metric perturbation (embedded in a $\unit[4]{s}$ segment) given binary parameters $\theta$.
Meanwhile, $\pi(\theta)$ is the prior distribution assumed for our analysis (described in Section~\ref{mdc}) and ${\cal N}$ is a normalization factor to ensure that $S_h(f)$ is a power spectral density with units of $\unit[]{Hz^{-1}}$.
In practice we carry out the integral numerically using $N$ Monte Carlo draws
\begin{align}
S_h(f) = \frac{1}{N} \sum_{k=1}^N \, 
\frac{|\tilde{h}_+(f|\theta_k)|^2 + |\tilde{h}_\times(f|\theta_k)|^2}{\cal N} .
\end{align}
Given a duty cycle $\xi$, the average power spectral density in a dataset with noise and signal is given by
\begin{align}
S_h'(f) = \xi \, S_h(f) .
\end{align}
The signal amplitude spectral density $[S_h(f)]^{1/2}$ is plotted in Fig.~\ref{fig:Omega}a (blue) alongside the noise amplitude spectral density (orange).

The  energy density associated with $S_h$ is
\begin{align}
\Omega_\text{gw}(f) = \frac{2\pi^2 f^3}{3H_0^2} S_h'(f) ,
\end{align}
where $H_0$ is the Hubble parameter.
In Fig.~\ref{fig:Omega}b, we plot $\Omega_{\mathrm{gw}}(f)$ for the Monte Carlo dataset used in our analysis given $\xi = \xitrue$ (solid blue).
Alongside, in thick black, we plot a power-law integrated sensitivity curve~\cite{locus}, which shows the $\unit[1]{\sigma}$ sensitivity of the cross-correlation search for the Hanford-Livingston network operating at design sensitivity for one year.

Since our dataset includes only binaries with luminosity distances $d_L\leq\unit[\dmaxlab]{Gpc}$, our dataset includes only a fraction of the full stochastic background.
Using Fig.~2c from~\cite{Callister}, we estimate that the full stochastic background has an energy density about three times larger than the stochastic background from events with $z<\zmax$.
The red dashed curve shows $\Omega_\text{gw}(f)$ scaled by a factor of three to roughly estimate the total gravitational-wave energy density including the contribution from high redshifts.
The dashed red line reaches a value of $3\times10^{-10}$ at $\unit[25]{Hz}$, which is just below the allowed range predicted in~\cite{GW170817_stoch}: $\Omega_\text{gw}(f=\unit[25]{Hz})=1.1^{+1.2}_{-0.7}\times10^{-9}$.
The fact that the dashed red line does not intersect the black sensitivity curve indicates that the background is not detectable with a year of data using cross-correlation.

The signal-to-noise ratio of the cross-correlation search is given by
\begin{align}
\rho_\text{cc} = \sqrt{2 n} \left[ \sum_k 
\frac{\Gamma_{12}^2(f_k)\, S_h^2(f_k)}{P_1(f_k) \, P_2(f_k)} \right]^{1/2} .
\end{align}
See Eq.~22 in~\cite{locus}.
The sum over $k$ ranges over frequency bins.
Here, $P_1, P_2$ are the noise power spectral densities of detectors $1$ and $2$.
The variable $\Gamma_{12}$ is the overlap reduction function given in Eq.~15 of~\cite{locus}; and derived in~\cite{christensen}.
Assuming one year of data at design sensitivity, we obtain $\rho_\text{cc}=0.15$ for the dataset used in our analysis, which is limited to redshifts $z<\zmax$.
Extrapolating to arbitrarily high redshifts, we estimate $\rho_\text{cc}=0.46$.

\end{appendix}

\bibliography{optimal}

\end{document}